\documentclass[11pt,a4paper]{article}
\usepackage{jcappub}
\usepackage{float}
\usepackage[english]{babel}
\usepackage[utf8]{inputenc}
\usepackage{pgfplots}
\pgfplotsset{compat=1.18}
\usepackage{cancel}
\usepackage{graphicx}
\usepackage{orcidlink} 
\usepackage{amsmath,amssymb,amsfonts,amsthm}
\usepackage{bm,bbm}
\usepackage{tikz,stackrel}
\usepackage{pstricks}
\usepackage{pifont} 
\usepackage{empheq}
\usepackage{graphicx}
\usepackage{subcaption}
\newcommand\nn{\nonumber\\}

\newcommand\bb{\bibitem}
\newcommand{\bc}{\begin{center}}
\newcommand{\ec}{\end{center}}
\newcommand{\be}{\begin{equation}}
\newcommand{\ee}{\end{equation}}
\newcommand{\ba}{\begin{eqnarray}}
\newcommand{\ea}{\end{eqnarray}}
\def\bs{\begin{subequations}}
\def\es{\end{subequations}}
\newcommand{\ben}{\begin{equation*}}
\newcommand{\een}{\end{equation*}}
\newcommand{\ban}{\begin{eqnarray*}}
\newcommand{\ean}{\end{eqnarray*}}
\renewcommand{\leq}{\leqslant}
\renewcommand{\geq}{\geqslant}
\def\a{\alpha}
\def\b{\beta}
\def\de{\delta}
\def\g{\gamma}
\def\la{\lambda}
\def\k{\kappa}

\def\Om{\Omega}
\def\om{\omega}

\def\G{\Gamma}
\def\t{\tau}  
\def\s{\sigma}

\def\N{\nabla}

\def\cF{\mathcal{F}}

\def\cK{\mathcal{K}}
\def\cL{\mathcal{L}}

\def\cO{\mathcal{O}}

\def\ds{d_\textsc{s}}

\def\p{\partial}

\def\B{\Box}
\def\lst{\ell_*}
\newcommand{\Eq}[1]{(\ref{#1})}
\newcommand{\Eqq}[1]{eq.~(\ref{#1})}
\newcommand{\Eqqs}[1]{eqs.~(\ref{#1})}

\def\cob{\color{blue}}

\def\sech{{\rm sech}}
\newcommand{\au}[2]{#1.~#2}
\newcommand{\book}[5]{\emph{#1}, #2, #3, #4 (#5)}
\newcommand{\books}[4]{\emph{#1}, #2, #3 (#4)}
\newcommand{\oarX}[1]{\href{http://arxiv.org/abs/#1}{{\ttfamily\cob #1}}}
\newcommand{\arX}[1]{\href{http://arxiv.org/abs/#1}{{\ttfamily\cob arXiv:#1}}}
\newcommand{\doin}[6]{\href{http://dx.doi.org/#1}{{\cob {\it #2 #3} {\bf #4} (#6) #5}}}
\newcommand{\doinn}[5]{\href{http://dx.doi.org/#1}{{\cob {\it #2} {\bf #3} (#5) #4}}}
\newcommand{\doij}[5]{\href{http://dx.doi.org/#1}{{\cob {\it #2} {\bf #3} (#5) #4}}}
\newcommand{\ndoin}[6]{\href{#1}{{\cob {\it #2 #3} {\bf #4} (#6) #5}}}

\newcommand{\procsinm}[5]{in \emph{#1}, #2 (eds.), #3, #4 (#5)}

\newcommand{\tia}[1]{\textit{#1},}
\newcommand{\dotBox}{%
  \mathop{%
    \ooalign{%
      $\Box$\cr
      \hfil\raise0.1ex\hbox{$\cdot$}\hfil\cr
    }%
  }%
}

\def\lp{\ell_{\rm Pl}}

\def\rme{e}
\def\rmd{d}
\def\rmi{i}
\def\Re{{\rm Re}}

\newcounter{listcounter}

\allowdisplaybreaks
\begin{document}

\renewcommand{\thefootnote}{\fnsymbol{footnote}}

\title{Cosmology of fractional gravity}

\author[a,b]{Iván Salvador-García\,\orcidlink{0009-0002-2408-9452}}
\affiliation[a]{Instituto de Estructura de la Materia, CSIC, Serrano 121, 28006 Madrid, Spain}
\affiliation[b]{Departamento de Física Teórica, Universidad Autónoma de Madrid, 28049 Madrid, Spain}
\emailAdd{ivan.salvador@iem.cfmac.csic.es}

\author[a,*]{and Gianluca Calcagni\,\orcidlink{0000-0003-2631-4588}\note{Corresponding author.}}
\emailAdd{g.calcagni@csic.es}

\abstract{This is a first study of the cosmology of classical fractional gravity, a nonlocal proposal endowed with self-adjoint fractional d'Alembertian operators which serves as the basis for an ultraviolet-complete theory of quantum gravity. We derive the classical covariant nonlocal equations of motion for an arbitrary fractional exponent $\gamma$ and reduce them to the Friedmann equations on a homogeneous and isotropic cosmological background. We find that de Sitter is an exact stable solution and that hyperbolic cosine bouncing exact solutions are sustained by phantom ($w<-1$) or ghost ($\rho<0$) fluids, in the latter case with a new type of finite-future singularity in the barotropic index. Working with a more general bouncing profile, we verify that the theory does support bounces where the null and weak energy conditions are respected for certain non-fine-tuned values of $\gamma$. On a branch of solutions, different representations of the form factor are equivalent, in line with the notion that fractional field theories are formulated on a universality class of form factors. We compare these preliminary results with what obtained in multi-fractional cosmological models mimicking the spacetime geometry of fractional quantum gravity.}


\maketitle
\renewcommand{\thefootnote}{\arabic{footnote}}


\section{Introduction}\label{intro}

The search for a quantum field theory (QFT) of gravity has delivered several theories with interesting features and many challenges. One of these approaches is perturbative string theory, which is based on QFT tools applied to extended objects; there, spacetime emerges at low energies and is of higher dimension \cite{string1,string2}. The notion of spacetime is also non-fundamental in loop quantum gravity, which is not a QFT and where basic degrees of freedom are not geometric \cite{loop1,thi01}. The mathematical and conceptual complexity of these and many other frameworks \cite{Bam24} is accompanied by a generalized difficulty to extract testable, falsifiable predictions.

Aside from these major candidates for a theory of quantum gravity, efforts in applying perturbative QFT to the gravitational interaction in four-dimensional spacetime have been resumed recently. Classic negative results
on the perturbative renormalizability of Einstein gravity \cite{tHooft:1974toh,Deser:1974cz,Deser:1974cy,Deser:1974xq,Goroff:1985sz,Goroff:1985th,vandeVen:1991gw,Bern:2015xsa,Bern:2017puu} and on the unitarity of its quadratic extensions \cite{Stelle:1976gc,Stelle:1977ry,Julve:1978xn,Fradkin:1981hx,Fradkin:1981iu,Avramidi:1985ki} had discouraged these lines. To get renormalizability, one includes higher-derivative curvature terms but these introduce higher-order derivatives which lead to a loss of unitarity at the quantum level. However, results in the last two decades have renewed hopes that gravity could be quantized in the same way as the other fundamental forces of Nature.
Proposals in this direction include nonlocal quantum gravity (NLQG) \cite{Kuzmin:1989sp,Tomboulis:1997gg,Modesto:2011kw,Biswas:2011ar,Biswas:2013cha,Modesto:2014lga,Modesto:2015lna,Dona:2015tra,Modesto:2016max,Calcagni:2018gke,Briscese:2018oyx,Buoninfante:2018mre,Calcagni:2023goc,Calcagni:2024xku,Briscese:2024tvc} (reviewed in \cite{Modesto:2017sdr,Buoninfante:2022ild,BasiBeneito:2022wux,Koshelev:2023elc}) and fractional quantum gravity (FQG), a special case of fractional QFT \cite{Calcagni:2024xku,Calcagni:2021ljs,Calcagni:2021ipd,Calcagni:2021aap,Calcagni:2022shb,Calcagni:2025wnn,Briscese:2026jcf,Calcagni:2026sfx,Anselmi:2026hvr}. Both  use standard perturbative QFT with a usual notion of Feynman diagrams, renormalizability, perturbative unitarity, and so on, all adapted to the presence of nonlocal operators. 

Cosmology is a great arena to test theories of modified and quantum gravity. The concordance $\Lambda$CDM model based on Einstein theory together with the mechanism of inflation have been successful in explaining observations such as the abundance of light elements or the properties of the cosmic microwave background (CMB) \cite{Planck:2018vyg,Akrami:2018odb}. In particular, Starobinsky's non-renormalizable extension $R+R^2$ of Einstein gravity \cite{Starobinsky:1980te} accounts for an inflationary scenario favoured by \textsc{Planck} observations. Despite the achievements of the $\Lambda$CDM model, it has been challenged by recent observations of baryon acoustic oscillations that indicate a preference for a dynamical dark-energy component \cite{DESI:2025zgx,Efstathiou:2024dvn,Chaudhary:2025uzr}. The $H_0$ and $S_8$ tensions \cite{Pantos:2026cxv,Pantos:2026koc} and the pressure suffered by the Starobinsky and $\a$-attractor inflationary models based on higher value of the scalar spectral index $n_{\rm s}$ \cite{AtacamaCosmologyTelescope:2025nti,Hazra:2024nav,Drees:2025ngb,Ferreira:2025lrd,McDonough:2025lzo} are opening a new season of precision cosmology where our established picture of the universe is being put into question. Theses discrepancies between different data-sets are of unclear origin and their interpretation still lies at the border between systematic effects and new physics. This stimulates a frank assessment of the intrinsic limitations of working with bottom-up models made with \emph{ad hoc} fields and assumptions. However, trying to describe the universe within a top-down theory of quantum gravity and matter is not straightforward and remains an open problem. For example, in relation with the early cosmic evolution, theoretical research concerning the existence of a non-singular beginning \cite{Biswas:2005qr,Biswas:2010zk,Koshelev:2012qn,Calcagni:2013vra,Koshelev:2014voa,Modesto:2022asj} and the embedding of Starobinsky-like scenarios \cite{Koshelev:2016xqb} has been done in NLQG but, unfortunately, with little on the side of new falsifiable predictions. As in many other cases in quantum gravity, the problem is not just to embed successful models into a more fundamental framework but to go beyond them and offer novel tests of such framework, as attempted in \cite{Calcagni:2022tuz}. 

These topics have not been tackled yet in FQG. A plethora of cosmological models have been explored in multi-scale or multi-fractal scenarios with non-trivial measures and ordinary, weighted or fractional derivatives \cite{Roberts:2009ix,Calcagni:2010bj,Shchigolev:2010vh,Karami:2012ra,Lemets:2012fp,Sheykhi:2013rla,Shchigolev:2013jq,Calcagni:2013yqa,Chattopadhyay:2013mwa,Shchigolev:2015rei,Rami:2015kha,Maity:2016dfv,Debnath:2019isy,Calcagni:2016ofu,Jawad:2016piq,Sadri:2018lzz,Das:2018bxc,Maity:2019qbv,Ghaffari:2019qcv,Jawad:2019xdv,Torres:2020xkw,Moniz:2020emn,Calcagni:2020ads,Mamon:2020ocb,Barrientos:2020kfp,Landim:2021www,Shchigolev:2021lbm,Jalalzadeh:2022uhl,Garcia-Aspeitia:2022uxz,Rasouli:2022bug,Micolta-Riascos:2023mqo,Junior:2023fwb,deOliveiraCosta:2023srx,Marroquin:2024ddg,Jalalzadeh:2024qej,Rasouli:2024crg,Micolta-Riascos:2024vbm,Kotal:2025hus,Rasouli:2025qix,Oliveira:2026hjx,Rasouli:2026hfy,Rasouli:2026vei,Chaturvedi:2026cvz}. These models have similarities with FQG due to the universality of the anomalous scaling of fractal-like integration measures and kinetic terms but none of them proceeds from a fundamental theory, since they cannot be rendered renormalizable and unitary at the same time \cite{Calcagni:2021ipd,Calcagni:2016azd}. Therefore, it may be misleading to extrapolate any result obtained in these models to FQG. The object of the present study is to begin to fill this gap. Although it is difficult to explore the space of solutions of the theory because of its nonlocality, we build the foundations for future work on the cosmology of FQG by finding the general equations of motion on a Friedmann--Lema\^itre--Robertson--Walker (FLRW) background. We then focus on the existence of non-singular and accelerating solutions. On one hand, it is relatively easy to obtain stable acceleration from curvature, de Sitter being the simplest example. On the other hand, a hyperbolic cosine and more general bouncing profiles seem to be unnaturally sustained by exotic matter and, for some sign choice of a parameter, characterized by a new type of sudden-future singularity in the equation of state, which we call zero-energy moment. However, there exist corners in the parameter space of the theory where bounces are sustained by ordinary matter, for certain values of $\g$ that do not require fine tuning.

The layout of this manuscript is the following. In section~\ref{fgra}, we recall the motivation of fractional QFT (section~\ref{sett}) and write down the action in fractional gravity, deriving the covariant modified Einstein equations for two different representations of the fractional power of the d'Alembertian (section~\ref{modeesec}). In section~\ref{friedsecgen}, we symmetry-reduce these equations of motion to an isotropic and homogeneous FLRW cosmology. Such equations are of infinite order and hence hard to solve. Thus, in section~\ref{generalansatzsec} we simplify the covariant equations of motion to the case of curvature tensors verifying a certain \emph{Ansatz} inspired by a similar approach employed in nonlocal cosmological models with entire form factors \cite{Biswas:2005qr}. In the simplified case of a purely $R$-dependent action, we check not only whether our theory admits non-singular bouncing solutions, but also whether different representations of the fractional d'Alembertian give the same solutions or not (the answer is Yes). A dynamical system analysis for the de Sitter exact solution is performed in section~\ref{dynsys}. Conclusions are in section~\ref{disc}, where we also compare our results with those of other multi-fractional models. The bulk of calculations is confined into three appendices.


\section{Equations of motion} \label{fgra}

In this section, after a very brief review of fractional QFT (section~\ref{sett}), we present the covariant modified Einstein equations (section~\ref{modeesec}) and their associated Friedmann equations (section~\ref{friedsecgen}). 


\subsection{Setting}\label{sett}

Different approaches to quantum gravity show a common feature, the change of spacetime dimension with the probed scale \cite{tHooft:1993dmi,Ambjorn:2005db,Lauscher:2005qz,Benedetti:2008gu,Modesto:2008jz,Horava:2009if,Calcagni:2009kc,Carlip:2009kf,Benedetti:2009ge,Modesto:2009qc,Calcagni:2013eua,Calcagni:2016edi,Carlip:2017eud,Mielczarek:2017cdp,Carlip:2019onx}. In some cases, this dimensional flow corresponds to a spacetime with multi-fractal geometry. Early attempts to define fractal spacetimes showed the difficulty of placing fields on a nowhere-differentiable fractal geometry \cite{Stillinger:1977mt,Svozil:1985ha,Eyink:1989dv,Eyink:1989se}. The use of
fractional calculus made the task easier because it approximates fractal geometry to a continuum \cite{Calcagni:2016azd,Calcagni:2021ipd}. Multi-fractional spacetimes are characterized both by a multi-fractional measure (which is trivial in some cases such as FQG) and by 
kinetic operators with anomalous scaling, which may contain derivatives of both integer and fractional order. The geometry of FQG is selected among other possibilities of the multi-fractional paradigm according to the following steps:
\begin{enumerate}
    \item Dimensional flow with a non-standard spacetime measure and ordinary second-order kinetic terms was found not to improve renormalizability \cite{Calcagni:2016azd}, which was actually the main goal. Therefore, the measure is kept trivial while forcing derivatives of non-integer order.
    \item Demanding diffeomorphism invariance in general and Lorentz invariance in local inertial frames, a fractional d'Alembertian is chosen instead of fractional derivatives \cite{Calcagni:2021ljs}. 
    \item The presence of a kinetic term with higher-order or non-integer derivatives can be understood as the spectral dimension $\ds$ of spacetime changing with the probed scale. In particular, $\ds$ depends on the fractional order $\g$ of the derivatives \cite{Calcagni:2012qn,Calcagni:2012rm}. The relation ranges from $\ds \simeq 4/\g$ in the ultraviolet (UV) to $\ds \approx 4$ in the infrared (IR). In this work, the separation between the IR and UV regimes will be set by a single length scale $\lst$, which in the case of a quantum theory of gravity is expected to be of order of the Planck scale $\lp$.
\end{enumerate}
For any given scaling, there are various ways to write a covariant fractional form factor \cite{Calcagni:2026sfx}. Here we employ the so-called hermitian simple form factor such that the action for gravity and matter is \cite{Calcagni:2026sfx}
\be\label{gravac}
S=\frac{1}{2\k^2}\int\rmd^4x\,\sqrt{|g|}\,\left[R+c_0\lst^2 R(\lst^2\dotBox)^{\g-2}\,R+c_2\lst^2 G_{\mu\nu}(\lst^2\dotBox)^{\g-2}\,R^{\mu\nu} \right]+S_{\rm m} \,,
\ee
with $\k^2 = 8 \pi G$, $G$ Newton's constant and $S_{\rm m}$ the matter action. The operator $\dotBox$ is self-adjoint and will be presented shortly. As discussed in \cite{Calcagni:2022shb,Calcagni:2026sfx}, depending on the value of $\g$ the theory is:
\begin{itemize}
    \item $\g <2$: non-renormalizable. There is an infinite number of divergences at all loop orders.
    \item $\g=2$ (Stelle gravity): strictly renormalizable: only a finite number of divergences appear but occur at all orders in perturbation theory.
    \item $\g>2$: super-renormalizable. Only a finite number of Feynman diagrams superficially diverge, up to some finite loop order.
\end{itemize}


\subsection{Modified Einstein equations} \label{modeesec}

We hereby obtain the equations of motion stemming from \eqref{gravac} for two representations of the self-adjoint operator $\dotBox$. Previous derivations \cite{Calcagni:2021aap,Calcagni:2022shb} were based upon the non-hermitian version of the theory ($\dotBox\to -\B$), which is unphysical since it does not have a well-defined classical limit. We work in $D$ topological dimensions and in signature $(-,+,\cdots,+)$.

\subsubsection{Self-adjoint Balakrishnan--Komatsu representation}

Let us start with the self-adjoint Balakrishnan--Komatsu representation of the fractional d'Alembertian \cite{Calcagni:2025wnn}: 
\be
{\dotBox}^{\g-2} = \frac{1}{\G(n-\g/2+1)} \int_0^{+\infty} \rmd \t \,  \t^{n-\g/2} (\B^2)^n e^{-\t \B^2} \, , \quad n < \text{Re} \,  \frac{\g}{2} < n+1 \in \mathbb{N} \, . \label{fracdalem2}
\ee
The modified Einstein equations stemming from \eqref{gravac} are (see appendices \ref{appA1} and \ref{appA2} for details and the definitions of $\vartheta_{\mu \nu} (R,R)$ and $\Theta_{\mu \nu} (R_{\a \b},G^{\a \b})$)
\be
\kappa^2 T_{\mu \nu} = \left[1+c_2 \lst^{2(\g-1)} {\dotBox}^{\g-2} \,  \B \right] G_{\mu \nu}+\lst^{2(\g-1)}\cO_{\mu\nu}\,,
\ee
where
\ba
\cO_{\mu \nu} &=& 2 c_0 \left[g_{\mu \nu} \B- \N_{(\mu} \N_{\nu)} \right] {\dotBox}^{\g-2} R \nonumber \\
&&+c_2 g_{\mu \nu} \N^\s \N^\t {\dotBox}^{\g-2} G_{\s \t}-2 c_2 \N^\s \N_{(\mu} {\dotBox}^{\g-2} G_{\nu) \s} \nonumber \\
&& + c_0 \left(G_{\mu \nu}+R_{\mu \nu} \right) {\dotBox}^{\g-2}R-\frac{1}{2} c_2 g_{\mu \nu} G_{\s \t} {\dotBox}^{\g-2} R^{\s \t} \nonumber \\
&& +2c_2 G_{(\mu}^\s {\dotBox}^{\g-2} G_{\nu) \s}+ \frac{c_2}{2}  \left(G_{\mu \nu} {\dotBox}^{\g-2}R+R {\dotBox}^{\g-2} G_{\mu \nu} \right) \nonumber \\
&& +c_0  \vartheta_{\mu \nu} \left(R,R \right)+c_2 \Theta_{\mu \nu} \big(R_{\a \b}, G^{\a \b} \big)\,. \label{gen00eq}
\ea
The trace of \eqref{gen00eq} is
\be
\kappa^2 \, T = \left(1-\frac{D}{2} \right) \left[1+c_2 \lst^{2(\g-1)} {\dotBox}^{\g-2} \B \right] R+\lst^{2(\g-1)}\cO\,,
\ee
where
\ba
\cO &=& 2 c_0 (D-1) \B {\dotBox}^{\g-2}R +(D-2) \, c_2 \N^\s \N^\t {\dotBox}^{\g-2} G_{\s \t}\nonumber \\
&& +\left(\frac{D}{4}-1 \right) \left[(c_2-2c_0)R {\dotBox}^{\g-2}R-2c_2 R_{\mu \s} {\dotBox}^{\g-2} R^{\mu \s} \right] \nonumber \\
&& +c_0 \vartheta_\mu^\mu \left(R,R \right)+c_2 \Theta_\mu^\mu \big(R_{\a \b}, G^{\a \b} \big) \,. \label{gentreq}
\ea
Note that the second line vanishes in $D=4$ dimensions.

\subsubsection{Self-adjoint Fresnel representation}

Using the Fresnel representation for the fractional d'Alembertian \cite{Calcagni:2025wnn} 
\ba
\hspace{-1cm}&&\B_{\text{F}}^{\g-2} = \frac{1}{\Gamma(n+2-\g) \cos{\frac{\pi(n+2-\g)}{2}}} \text{Re} \int_0^\infty \rmd \t \t^{n+1-\g} \,  \B^n \, \rme^{\rmi \t \B} \,,\nn
\hspace{-1cm}&&\qquad\qquad\qquad\qquad\qquad\qquad\qquad\qquad\qquad\qquad\qquad\qquad n+1 < \text{Re} \,  \g < n+2 \in \mathbb{N} \,, \label{fracdalem4}
\ea
the equations of motion are the same as \eqref{gen00eq}--\eqref{gentreq} upon replacing $\dotBox\to\B_\text{F}$, $\Theta_{\mu \nu}(R_{\a \b},G^{\a \b})\to\Xi_{\mu \nu}(R_{\a \b},G^{\a \b})$ and $\vartheta_{\mu \nu}(R_{\a \b},G^{\a \b})\to\xi_{\mu \nu}(R_{\a \b},G^{\a \b})$, where the definitions of $\xi_{\mu \nu}(R,R)$ and $\Xi_{\mu \nu}(R_{\a \b},G^{\a \b})$ are given in appendix~\ref{appA3}. 


\subsection{Modified Friedmann equations} \label{friedsecgen}

Having obtained the covariant equations of motion, in this section we work out a symmetry reduction of \eqref{gen00eq}--\eqref{gentreq} on a $D=4$ flat FLRW cosmological background with line element
\be 
\rmd s^2 = -\rmd t^2+a^2(t)\sum_{i=1}^{3}\rmd x_i^2\,, \label{friedmet}
\ee
where $t=x^0$ is proper time. Regarding the energy-momentum content, we take a perfect fluid with energy density $\rho=T_{00}$ and pressure $p=T_i^{i}/3$:
\be\label{peflu}
T_{\mu\nu}=(\rho+p)\,u_\mu u_\nu+p\,g_{\mu\nu}\,,
\ee
where $u^\mu=\rmd x^\mu/\rmd t$ is the comoving $4$-velocity tangent to a fluid element world-line ($u_\mu u^\mu=-1$). Equations below use the Hubble parameter $H \coloneqq \dot{a}/a$. 

Thanks to the conservation $\N^\mu T_{\mu\nu}=0$ of the energy-momentum tensor in a general nonlocal covariant quadratic action \cite{Biswas:2013cha}, the perfect fluid obeys the usual conservation law
\be
\dot{\rho}+3H(\rho+p) = 0\,,
\ee
which can also be obtained by combining the Friedmann equations together:
\ba 
\hspace{-1.2cm}\k^2 \rho &=&  3 H^2 + c_2 \lst^{2(\g-1)} {\dotBox}^{\g-2} \left[-\frac{\ddot{Q}}{2} + \frac{3}{2} H \dot{Q} -(\ddot{P}+3H \dot{P})\right]  \nonumber \\ &&   -c_2 \lst^{2(\g-1)} \left[\N^\s \N^\t {\dotBox}^{\g-2} \left(Q u_\s u_\t \right)+2 \N^\s \N_t {\dotBox}^{\g-2} \left(Q u_\s \right) \right] \nonumber \\ 
&& +\lst^{2(\g-1)} \bigg\{6 \, c_0 \, Q -24 c_0 H \p_t   + \, c_2  \left(\p_t^2+3H \p_t \right)\bigg\} {\dotBox}^{\g-2} P \nonumber \\
&& +\lst^{2(\g-1)} \bigg\{-\frac{3 \, c_0}{2} Q-\frac{3 \, c_2}{4} Q  +6 c_0 H \p_t   -\frac{c_2}{2} \left(\p_t^2+3H \p_t \right)\bigg\} {\dotBox}^{\g-2} Q \nonumber \\ 
&& + c_0 \lst^{2(\g-1)} \vartheta_{00} \left(R,R \right)+c_2 \lst^{2(\g-1)}\Theta_{00} \big(R_{\a \b}, G^{\a \b} \big) \,, \label{fried1}
\ea
plus the trace equation
\ba 
\k^2 (3p-\rho) &=& -6 (2H^2+\dot{H} ) + \lst^{2(\g-1)} (6 \, c_0 - c_2) {\dotBox}^{\g-2} 
\left[-(4 \ddot{P}-\ddot{Q})-3H (4 \dot{P}-\dot{Q})\right] \nonumber \\
&& +  2 \, c_2 \lst^{2(\g-1)} (-\p_t^2 + 3H \p_t) {\dotBox}^{\g-2} 
\left(\frac{Q}{2}-P \right) + 2 \, c_2 \lst^{2(\g-1)} \N^\s \N^\t {\dotBox}^{\g-2} (Q u_\s u_\t)  \nonumber \\
&& + c_0 \lst^{2(\g-1)} \vartheta_\mu^\mu \left(R,R \right)+c_2 \lst^{2(\g-1)}\Theta_\mu^\mu \big(R_{\a \b}, G^{\a \b} \big)\,, \nonumber \\  \label{fried2}
\ea
where \cite{Gurses:2020kpv,Gurses:2024tka}
\ba
&& P \coloneqq 3H^2 + \dot{H} \,,\qquad Q \coloneqq -2\dot{H}\,,\\
&& R_{00}= Q-P\,,\qquad R_{ij}=Pa^2\de_{ij}\,,\qquad R=6(2H^2+\dot H)=4P-Q\,.
\ea
Details of the calculation are given in appendix~\ref{appB}. Equations \Eq{fried1} and \Eq{fried2} are extremely complicated and, in fact, we have not made most of the terms explicit.


\section{Bouncing toy models} \label{generalansatzsec}

As we have seen, nonlocal gravity field equations contain infinite derivatives. This makes the search of classical solutions a hard problem that can be tackled with the diffusion method \cite{Calcagni:2025wnn}. Here, however, we use a more limited but simpler approach. In previous works \cite{Biswas:2005qr,Biswas:2010zk,Koshelev:2014voa,Koshelev:2016xqb,Koshelev:2012qn,Biswas:2012bp}, some progress was obtained by restricting the solutions of toy models inspired by NLQG to a certain \emph{Ansatz} which reduces the nonlocal equations to finite order. We start by introducing the \emph{Ansatz} in section~\ref{ouransatzsec} and obtain the reduced modified Einstein and Friedmann equations in $D$ dimensions for both representations \Eq{fracdalem2} and \Eq{fracdalem4}. In section~\ref{cosmosolsec}, we explore bouncing solutions.


\subsection{Ansatz} \label{ouransatzsec}

In order to simplify the full equations of motion, in the case $c_2 \neq 0$ we would need an \emph{Ansatz} for a rank-2 tensor as
\be\label{ansatzricci}
\B R_{\mu \nu} = \la_1 R_{\mu \nu}+\la_2 g_{\mu \nu} \,,
\ee
where $\la_1$ and $\la_2$ are constants. This is a stronger \emph{Ansatz} than the usual one on the Ricci scalar \cite{Biswas:2005qr,Biswas:2010zk,Koshelev:2014voa,Koshelev:2016xqb,Koshelev:2012qn,Biswas:2012bp}:
\be
\B R = \la_1 R+\la_2 D\,. \label{ansatzR} 
\ee
From here on, we stick with \eqref{ansatzR} and study only the case $c_2 = 0$ in \eqref{gravac}. This simplified setting still gives precious information on the existence and behaviour of bouncing solutions.

For $1\leq n \in \mathbb{N}$, \eqref{ansatzR} implies
\be
\Box^n R = \lambda_1^n  \left(R + \frac{\la_2}{\la_1} D \right).
\ee
Given an analytic function $f(z) = \sum_n f_n z^n$, it is also useful to compute
\ba
f(\Box) R = f(\la_1)R+\frac{\la_2}{\la_1} \left[f(\la_1)-f_0 \right]D \label{fR} \,.
\ea
In particular, the action of the fractional d'Alembertian in the Balakrishnan--Komatsu representation \eqref{fracdalem2} is
\be
{\dotBox}^\g R = |\la_1|^\g \left(R+\frac{\la_2}{\la_1}D \right).
\ee
Without loss of generality, we set $n=1$ in \eqref{fracdalem2}, so that we consider the range $2<\g<4$ which includes the unitarity$+$renormalizability range $2<\g<3$ for the hermitian simple theory \cite{Calcagni:2026sfx}. Therefore, for $c_2 = 0$ the Einstein equations become (appendix~\ref{appC})
\ba
\k^2 T_{\mu \nu} &=& G_{\mu \nu}+2 c_0 \lst^2 (|\la_1| \lst^2)^{\g-2} \left[\la_1 \left(R+\frac{\la_2}{\la_1} D\right) g_{\mu \nu} -\N_\mu \N_\nu R  \right] \nonumber \\
&&+c_0 \lst^2 (|\la_1| \lst^2)^{\g-2} \left(G_{\mu \nu}+R_{\mu \nu} \right) \left(R+\frac{\la_2}{\la_1} D\right) \nonumber \\
&&+ \frac{c_0 \lst^2 (|\la_1| \lst^2)^{\g-2}}{\la_1} \left[ (\g-2) \bar{\vartheta}_{\mu \nu} \left(R,R \right) +\frac{D}{2} (\g-3) \la_2 \left(R+\frac{\la_2}{\la_1}D \right) g_{\mu \nu}\right], 
\label{EinAns00}
\ea
and their trace is
\ba
\k^2 T &=& \left(1-\frac{D}{2} \right)R+c_0 \lst^2 (|\lambda_1| \lst^2){}^{\g-2} \left(R+\frac{\la_2}{\la_1} D\right)  \left[ 2(D-1) \la_1  + \left(2-\frac{D}{2} \right)R \right] \nonumber \\
&& +\frac{c_0 \lst^2 (\lst^2 |\la_1|)^{\g-2}}{\la_1}  \left[(\g-2) \bar{\vartheta}_\mu^\mu(R,R)+\frac{D^2}{2} (\g-3) \la_2 \left(R+\frac{\la_2}{\la_1} D \right) \right] \, . \label{EinAnsTr}
\ea
On a $D=4$ flat FLRW background, \Eqqs{EinAns00} and \eqref{EinAnsTr} become
\ba
\k^2 \rho &=& 3H^2+\frac{c_0 \lst^2 (|\la_1| \lst^2)^{\g-2}}{\la_1^2} \Big(-8 \la_2 \left[\la_1^2+(\g-3) \la_2 \right]
+6 \la_1 \Big\{-4 \left[\la_1^2+(\g-3) \la_2 \right]H^2\nn
&& -2 \left[\la_1^2 +(\g-2) \la_2 +3 \la_1 H^2 -36 (\g-2)H^4\right] \dot{H}+12 (\g-2) \dot{H}^3 \nn
&& + \left[-11 \la_1+12(\g-2)H^2 \right] \dot{H}^2- \left[8 \la_1-42(\g-2)H^2+3(\g-2) \dot{H} \right] H\ddot{H} \nn
&& +3(\g-2) \ddot{H}^2-\left[2 \la_1-6(\g-2)H^2-3(\g-2) \dot{H} \right] \dddot{H}  \Big\}\Big), \label{fried12}\\
\k^2 (3p-\rho) &=& -6\left(2H^2+\dot{H} \right)+\frac{4 c_0 \lst^2 (|\la_1| \lst^2)^{\g-2}}{\la_1^2} \Big(6 \la_1^2 \la_2+8(\g-3) \la_2^2 \nn
&& +3 \la_1 \Big\{-144(\g-2)H^4 \dot{H}-24(\g-2)\dot{H}^3   -84 (\g-2)H^3 \ddot{H} -66 (\g-2)H \dot{H} \ddot{H}  \nn
&& -3(\g-2) \ddot{H}^2 +\left[ 3 \la_1^2+4(\g-3) \la_2-6(\g-2) \dddot{H} \right]\dot{H}  \nn
&&  +2 \left[3 \la_1^2+4(\g-3) \la_2-84 (\g-2) \dot{H}^2-6(\g-2) \dddot{H} \right]H^2 \Big\}\Big), \label{fried22}
\ea
where we used the 00 component and the trace of \eqref{barvartheta}:
\ba
\bar{\vartheta}_{00}(R,R) &=& -\frac{1}{2}(\dot{R}^2+R \B R)\nonumber \\
&=& 18  \left[24 H^4 \dot{H}+4 \dot{H}^3+14 H^3 \ddot{H}-H \dot{H}\ddot{H}- \ddot{H}^2+ \dot{H} \dddot{H}+2H^2 (2 \dot{H}^2+ \dddot{H}) \right],\\
\bar{\vartheta}_\mu^\mu (R,R) &=& \frac{D}{2} R \B R +\left(1-\frac{D}{2} \right)\dot{R}^2 \nonumber \\
&=& -36 \left[48H^4 \dot{H}+8 \dot{H}^3+28 H^3 \ddot{H}+22 H \dot{H} \ddot{H}+\ddot{H}^2+2 \dot{H} \dddot{H}+4H^2(14 \dot{H}^2+\dddot{H}) \right].\nonumber \\
\ea

For the Fresnel representation \Eq{fracdalem4} with $n=1$, we are exactly in the unitarity$+$renormal\-izability range $2<\g<3$ of the hermitian simple theory. Then, for $c_2=0$ we get
\ban
\xi_{\mu \nu}^{(1)}(R,R) &=& |\la_1|^{\g-3} \bar{\vartheta}_{\mu \nu}(R,R)\,,\\
\xi_{\mu \nu}^{(2)}(R,R) &=& (\g-3) |\la_1|^{\g-3} \left[\bar{\vartheta}_{\mu \nu}(R,R)+D\frac{\la_2}{\la_1} \bar{\vartheta}_{\mu \nu}(1,R) \right],
\ean
so that
\ba
\k^2 T_{\mu \nu} &=& G_{\mu \nu}+2 \, c_0 \, \text{sgn}{(\la_1)}\lst^2 (|\la_1| \lst^2)^{\g-2} \left[\la_1 \left(R+\frac{\la_2}{\la_1} D\right) g_{\mu \nu} -\N_\mu \N_\nu R  \right] \nonumber \\
&&+c_0\, \text{sgn}{(\la_1)} \lst^2 (|\la_1| \lst^2)^{\g-2} \left(G_{\mu \nu}+R_{\mu \nu} \right) \left(R+\frac{\la_2}{\la_1} D\right) \nonumber \\ \label{freseom1}
&&+ \frac{c_0 \ell_*^2 (|\la_1| \ell_*^2)^{\g-2}}{\la_1} \text{sgn}(\la_1) \left[ (\g-2) \bar{\vartheta}_{\mu \nu}(R,R)+ \frac{D}{2} (\g-3) \la_2 \left(R+\frac{\la_2}{\la_1} D \right) g_{\mu \nu} \right ], \nonumber \\ \\
\k^2 T &=& \left(1-\frac{D}{2} \right)R+c_0\, \text{sgn}(\la_1) \lst^2 (|\lambda_1| \lst^2){}^{\g-2} \left(R+\frac{\la_2}{\la_1} D\right)  \left[ 2(D-1) \la_1  + \left(2-\frac{D}{2} \right)R \right] \nonumber \\
&&+ \frac{c_0 \ell_*^2 (|\la_1| \ell_*^2)^{\g-2}}{\la_1} \text{sgn}(\la_1) \left[ (\g-2) \bar{\vartheta}_\mu^\mu(R,R)+ \frac{D^2}{2} (\g-3) \la_2 \left(R+\frac{\la_2}{\la_1} D \right) \right] \, . \nonumber \\
\label{freseom2}
\ea
We can now compare the equations of motion for the two representations using the \emph{Ansatz} \Eq{ansatzR} (\Eqqs{EinAns00}--\eqref{EinAnsTr} for Balakrishnan--Komatsu and \Eqqs{freseom1}--\eqref{freseom2} for Fresnel) and see that they are identical except for a factor $\text{sgn}(\la_1)$ in the $c_0$ terms in the Fresnel case. This is because here we have considered a factor $\Box^n$ in \eqref{fracdalem4} instead of $|\Box|^n$ (see \cite{Calcagni:2025wnn}), which allows one to compute the variation of the action without having to deal with $\delta |\Box|^n$ factors. Thus, when $\la_1\geq 0$ the equations of motion with the \emph{Ansatz} \Eq{ansatzR} are equivalent in the two representations.


\subsection{Hyperbolic cosine bounce} \label{cosmosolsec}

Here we consider regular cosmological solutions with a hyperbolic cosine scale factor:
\be\label{bou1}
a(t) = \cosh\sqrt{\frac{\om}{2}}t\,,\qquad \om>0\,,
\ee
which was discussed in \cite{Biswas:2005qr} for nonlocal toy models with Lagrangian $\cL=R+R \cF_R(\Box)R$ and an analytic form factor $\cF_R(\Box)$. By imposing cancellation of some terms in the equations of motion, as well as radiation with positive energy density, the authors constrained this class of models. In the fractional case, we have a fixed $\cF_{R,2}(\Box)$ and only $c_0$, $c_2$ and $\g$ as free parameters. Let us study whether we can have the hyperbolic bounce \Eq{bou1} as an exact solution of \eqref{gravac}.

First, \Eq{bou1} verifies our \emph{Ansatz} only in its scalar form \eqref{ansatzR}, with $\la_1 = \om$ and $\la_2 = -3\om^2/2$. Therefore, we need to consider a reduced version of \eqref{gravac} with $c_2 = 0$. This originates a non-renormalizable toy model that, however, is expected to retain about the same properties of the full theory. Notice, in fact, that the $c_0$ term in the action \eqref{gravac} is nonlocal and we are still capturing non-trivial nonlocal effects. We start studying the case $\g>2$, interesting for power-counting renormalizability of the full theory. In particular, take $2<\g<4$. With the Balakrishnan--Komatsu representation, the Friedmann equations \eqref{fried12} and \eqref{fried22} become
\ba
\hspace{-1cm}\kappa^2\rho &=& \frac{3}{2}\om 
-\frac32\left[\om -\frac{18c_0}{\lst^2} (\lst^2 \om)^\g \right] \, \sech^2 \sqrt{\frac{\om}{2}}t \nonumber \\
\hspace{-1cm}&& - \frac{27c_0}{2 \lst^2} (\g-1) (\lst^2 \om)^\g \sech^4\sqrt{\frac{\om}{2}}t+\frac{9c_0}{\lst^2}(\g-2) (\lst^2 \om)^\g \sech^6\sqrt{\frac{\om}{2}}t\,, \label{00bounce1}\\
\hspace{-1cm}\kappa^2(3p-\rho) &=& -6\om +3 \left[\om-\frac{18c_0}{\lst^2} (\lst^2 \om)^\g \right] \text{sech}^2\sqrt{\frac{\om}{2}}t+\frac{18c_0}{\lst^2}(\g-2) (\lst^2 \om)^\g \text{sech}^6 \sqrt{\frac{\om}{2}}t \,.\label{tracebounce1}
\ea
Using the Fresnel representation of the form factor, we get exactly the same result because, as we saw above, the Friedmann equations using the \emph{Ansatz} are the same up to factors $\text{sgn}\,\la_1=1$. This is the first example of solution of fractional gravity on which the equivalence of the two representations \Eq{fracdalem2} and \Eq{fracdalem4} in the sense of belonging to the same universality class of operators is established.

Trying to match each term on the right-hand side of \Eqqs{00bounce1} and \Eq{tracebounce1} with a specific energy-momentum contribution on the left-hand side highlights the presence of exotic components, as follows. The time-independent term in \Eq{tracebounce1} can be produced by a cosmological constant with $(3p_\Lambda-\rho_\Lambda)\kappa^2 = -4 \rho_\Lambda \kappa^2= -4 \Lambda $, so that
\be
\Lambda = \frac{3}{2}\om>0 \, . \label{lamsol}
\ee
Looking for an exact solution with a minimal set of energy-momentum components, we note that the $\sech^2$ term can be reproduced by a cosmological fluid with equation of state $w = -1/3$, i.e.,  a frustrated network of cosmic strings \cite{Bucher:1998mh} with $\rho\sim a^{-2}$. However, here we do not invoke this component and rather note that the $\sech^2$ term is the only one that can be made vanish exactly when $\g$ is non-integer. Its cancellation in \Eq{tracebounce1} yields a relation between the characteristic scale of the bounce $\om$ and the parameters of the model:
\be
\om = (18 c_0)^{\frac{1}{1-\g}} \lst^{-2}  \,, \label{wsol}
\ee
so that we need $c_0>0$. We are left with the $\sech^6$ term and here we recognize a major difference between our framework and the one in \cite{Biswas:2005qr}. There, the action was tailored \emph{ad hoc} such that the $\sech^2$ and $\sech^6$ terms canceled and the only matter content was a cosmological constant $\Lambda$ plus radiation. In our case, we work with a fixed action and we do not enjoy such freedom. Indeed, we cannot cancel the $\sech^6$ term for non-integer $\g$ and we need to introduce stiff matter with $p= \rho$, such that its energy density scales as $\rho \sim a^{-6}$. An example of stiff matter is a free scalar field, for which
\be\label{stifene}
\rho_\phi = \rho_{\phi 0}\,\sech^6 \sqrt{\frac{\om}{2}}t\,,\qquad \rho_{\phi 0}=\left(\frac{\g}{2}-1 \right)(18 \, c_0)^{\frac{1}{1-\g}} (\lst\k)^{-2}\,.
\ee
Note that, as we are in the $2<\g<4$ range, the stiff matter energy density \eqref{stifene} is positive. 

The order zero, two and six in $\sech$ in \Eq{00bounce1} are already satisfied if we impose \eqref{lamsol}--\eqref{stifene}. We are left with the $\sech^4$ term, corresponding to radiation with energy density $\rho_{\rm r} \sim a^{-4}$ (this does not appear in the trace equations as radiation is traceless). Then
\ben
\rho_{\rm r} = \rho_{{\rm r} 0}\,\sech^4 \sqrt{\frac{\om}{2}}t\,,\qquad \rho_{{\rm r}0}=
\frac{3}{4}(1-\g) (18 c_0)^{\frac{1}{1-\g}}(\lst\k)^{-2}\,,
\een
which is negative since $\g>1$. Considering the ratio between the absolute value of radiation and stiff matter energy at $t=0$,
\ben
\frac{|\rho_{{\rm r}0}|}{\rho_{\phi 0}} = \frac{3(\g-1)}{2(\g-2)}>1  \, ,
\een
implying that the ghost radiation component dominates over stiff matter at the bounce. Computing the ratio of the cosmological constant energy density to radiation, for $2<\g<3$ one has
\ben
\frac{\rho_\Lambda}{|\rho_{{\rm r}0}|}=  \frac{2}{\g-1}>1 \,,
\een
so that the cosmological constant dominates at the bounce over all the other contributions, while for $\g>3$ radiation dominates.

\begin{figure}[ht]
	\bc
	\includegraphics[width=12cm]{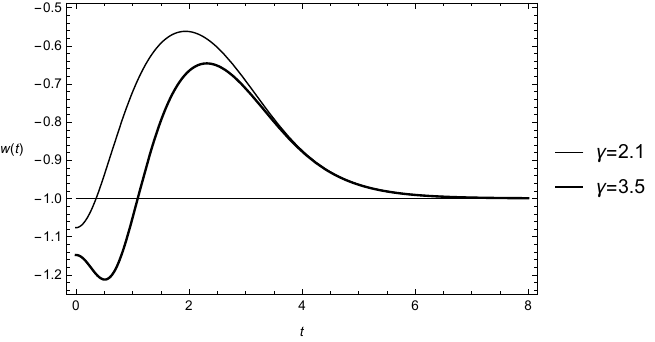}\\
    \vspace{.5cm}
    \includegraphics[width=12cm]{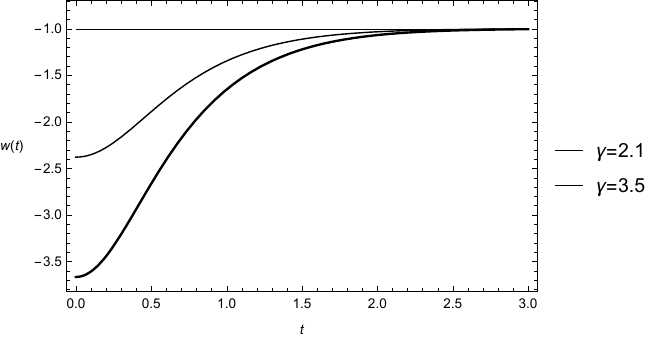}
	\ec
	\caption{\label{fig1} Examples of effective equation of state \Eq{eos} for $\om=1=\lst$ and $c_0=1$ (top) and $c_0=1/18$ (bottom). The $w=-1$ line is shown for reference. The curves are symmetric under time reflection $t\leftrightarrow -t$.}
\end{figure}

\begin{figure}[ht]
	\bc
	\includegraphics[width=12cm]{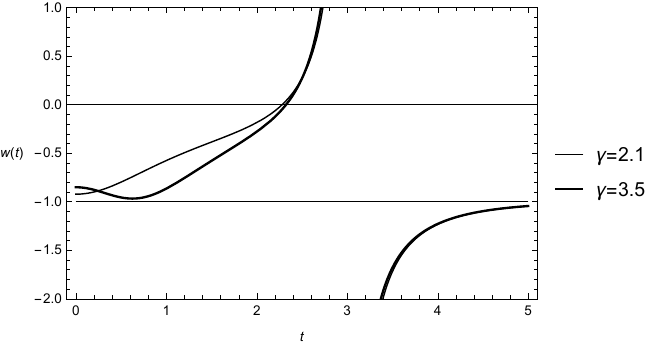}
	\ec
	\caption{\label{fig2} Examples of effective equation of state \Eq{eos} for $\om=1=\lst$ and $c_0=-1$. The $w=-1$ line is shown for reference. The curves are symmetric under time reflection $t\leftrightarrow -t$.}
\end{figure}

In order to gain more insight in the physics of bouncing cosmology in fractional gravity, we now proceed in a different way. Rather than separating matter contributions, we consider the global equation of state $w(t) \coloneqq p(t)/\rho(t)$:
\be
w(t) = -1-\frac{2}{3}\frac{(1-18y)\,\sech^2\sqrt{\frac{\om}{2}}t+18(\g-1)y\,\sech^4\sqrt{\frac{\om}{2}}t-18(\g-2)y\,\sech^6\sqrt{\frac{\om}{2}}t}{1-(1-18y)\,\sech^2\sqrt{\frac{\om}{2}}t-9(\g-1)y\,\sech^4\sqrt{\frac{\om}{2}}t+6(\g-2)y\,\sech^6\sqrt{\frac{\om}{2}}t}\,,\label{eos}
\ee
where $y\coloneqq c_0(\lst^2 \om)^{\g-1}$. Note that the $\sech^2$ terms disappear for the critical value $y=1/18$. At late times, $w\to -1$, while near the bounce
\be
w= -1 -\frac{2}{9(5-\g)y}+O(t^2)\,. 
\ee
Therefore, if $c_0>0$ ($y>0$) one always has a phantom-like equation of state with $w(0)<-1$, while if $c_0<0$ ($y<0$) one has an ordinary barotropic index $w(0)>-1$ that can be tuned to any desired value by choosing $\om$.
 The effective barotropic index $w(t)$ for $c_0>0$ and $c_0<0$ is plotted for positive times in figures~\ref{fig1} and \ref{fig2}, respectively.

When $c_0$ is positive and $y\neq 1/18$ (top panel of figure~\ref{fig1}), at the bounce the universe is dominated by a phantom component $w(0)<-1$ \cite{Caldwell:1999ew}, which then evolves through some transient regimes where $w$ stays negative and asymptotically reaches a de Sitter regime from above $w\gtrsim -1$ in the future. Since the bounce is symmetric, this implies that the universe began in the infinite past in a contracting accelerating phase. For the critical value $y=1/18$ (bottom panel of figure~\ref{fig1}), the barotropic index evolves monotonically from a strongly phantom regime $w(0)\ll -1$ at the bounce to de Sitter.

When $c_0<0$, the scenario changes considerably (figure~\ref{fig2}). At the bounce, the universe is filled with exotic matter with an apparently normal barotropic index $w(0)>-1$ but negative energy density and pressure. At some finite time $t=t_0$, the universe passes through a state where the energy-momentum components compensate one another and the energy density vanishes, $\rho(t_0)=0$, the barotropic index diverges but the scale factor and its derivatives are regular:
\be\label{sudd}
\rho(t_0)=0\,,\qquad w(t_0)=\infty\,,\qquad a(t_0)=a_{\rm s}\,,
\ee
where $a_{\rm s}$ is finite. After this zero-energy moment (ZEM), the cosmic expansion resumes with a positive energy density and tends asymptotically to de Sitter, $a(t)\simeq \exp(\sqrt{\om/2}\,t)/2$. The pre-bounce evolution is the mirror-symmetric of the above.

The singularity \Eq{sudd} of the barotropic index does not correspond to any of the sudden future singularities classified in \cite{Nojiri:2005sx}. Type I (big rip \cite{McInnes:2001zw,Briscese:2006xu}) is both a metric and matter singularity where $(a,\rho,p)\to (\infty,\infty,\infty)$ at some finite time $t_0$, with $w<-1$. Type II (Barrow singularity \cite{Barrow:2004xh,Barrow:2004hk}) is a pressure singularity such that $(a,\rho,p)\to (a_{\rm s},\rho_{\rm s},\infty)$, where quantities with a subscript s are non-zero and finite. In this case, $w(t_0)=\infty$ like the ZEM \Eq{sudd} but with a diverging pressure instead of a vanishing energy density. Type III \cite{Nojiri:2004pf,Stefancic:2004kb} is a matter singularity with $(a,\rho,p)\to (a_{\rm s},\infty,\infty)$ and $w(t_0)=\infty$; here both the energy density and pressure diverge. Type IV \cite{Nojiri:2005sx} is a metric singularity with $(a,\rho,p)\to (a_{\rm s},0,0)$ where higher derivatives of $H$ diverge but it is also a barotropic-index singularity since $w(t_0)=\infty$. However, here also the pressure vanishes.

In contrast, the ZEM \Eq{sudd} has $(a,\rho,p)\to (a_{\rm s},0,p_{\rm s})$. It is neither a metric nor a matter singularity but a barotropic-index singularity, where the universe momentarily enters a state of zero total energy at some point before and after the bounce.

To summarize the physical picture emerging from this analysis, hyperbolic cosine bouncing solutions in fractional gravity represent a non-singular universe plagued by highly exotic and possibly problematic components around the bounce, either of phantom ($w<-1$, $c_0>0$) or of ghost type ($\rho<0$, $c_0<0$).


\subsection{General bounce}\label{genbou}

The results of the previous sub-section indicate that, in the presence of fractional operators, the hyperbolic cosine bounce \Eq{bou1} is sustained by exotic matter that violates either or both the null energy condition (NEC)
\be\label{nec}
\textrm{NEC:}\qquad \rho(1+w) \geq 0\,,
\ee
and the weak energy condition (WEC) 
\be\label{wec}
\textrm{WEC:}\qquad \rho(1+w) \geq 0\,,\quad \rho \geq 0\,.
\ee
Since one would have expected that the constructive role of fractional operators in the quantum theory had corresponding benefits also in the classical one, it is important to understand whether the above negative outcome is typical of the theory or is a feature of the specific bouncing profile \Eq{bou1}. 
As a preliminary check, one can consider the sign of $\rho$ at the bounce coming from the Friedmann equation \Eq{fried12}. At the bounce time $t=0$, we have $H(0)=0$ and, by continuity a little after, $\dot H(0^+)>0$. Then, \Eqq{fried12} becomes
\ba
\k^2 \rho(0) &=& \frac{c_0 \lst^2 (|\la_1| \lst^2)^{\g-2}}{\la_1^2} \Big\{-8 \la_2 \left[\la_1^2+(\g-3) \la_2 \right]\nonumber \\
&& +6 \la_1 \Big[\la_1\left(-2\la_1\dot{H}-11 \dot{H}^2-2 \dddot{H}\right) \nonumber \\
&& +(\g-2)\left(-2\la_2\dot{H}+12 \dot{H}^3+3\ddot{H}^2+3\dot{H} \dddot{H}\right)   \Big]\Big\}.\label{fried12n}
\ea
In the case of local Starobinsky gravity ($\g=2$), the \emph{Ansatz} \Eq{ansatzR} holds in particular for $\la_2=0$ and radiation as the only energy-momentum component (appendix~\ref{appD2}). Then, only the second line of \Eqq{fried12n} survives and is reduced to \Eqq{starorho0}, which violates the WEC \Eq{wec} if we assume $c_0\lst^2=c>0$. In contrast, when $\g>2$ as in FQG there are two positive terms $O(\dot H^3)$ and $O(\ddot H^2)$ in the last line of \Eq{fried12n} that can compensate, in principle, the others to get a positive $\rho$.

In order to get a more rigorous and clear picture, we check whether the pathological matter content found in section~\ref{cosmosolsec} can be avoided in a  general bouncing profile. To pursue this goal, we follow the approach in \cite{Biswas:2010zk} and obtain four-dimensional bouncing cosmologies not restricted to the hyperbolic cosine. Let us work with an odd power series for the Hubble parameter (so that the bounce is symmetric under time reversal):
\be
H(t) = \lst^{-1} \sum_{m=0}^\infty a_{2m+1} \left(\frac{t}{\lst} \right)^{2m+1}. \label{Hseries}
\ee
One can compute the Ricci scalar as well as the d'Alembertian of the Ricci scalar:
\be
R = 6 \,  \lst^{-2} \sum_{m=0}^\infty A_{2m} \left(\frac{t}{\lst} \right)^{2m},
\ee
with
\ban
A_0 &=& a_1 \, , \\
A_2 &=& 2a_1^2+3a_3 \, , \\
A_4 &=& 4 a_1 a_3+5 a_5 \, , \\
A_6 &=& 2 a_3^2+4 a_1 a_5+7 a_7 \, , \\
&& \cdots \nonumber
\ean
and
\be
\B R = -12 \,  \lst^{-4} \sum_{m=0}^\infty B_{2m} \left(\frac{t}{\lst} \right)^{2m} \, ,
\ee
with
\ban
B_0 &=& 2 a_1^2+3a_3\, , \\
B_2 &=& 3 \left(2a_1^3+11a_1 a_3+10a_5 \right), \\
B_4 &=& 3 \left(10 a_1^2 a_3+13 a_3^2+30 a_1 a_5+35 a_7\right), \\
B_6 &=& 42 a_1^2 a_5 +151 a_3 a_5+7 a_1 (6a_3^2+25 a_7)\,. \\
&& \cdots 
\ean
In order to study these bounces in fractional cosmology, we ask them to verify the scalar \emph{Ansatz} \eqref{ansatzR}. By imposing it order by order in $t$, we obtain
\ban
a_3 &=& -\frac{1}{3} \left(2a_1^2+\frac{1}{2} a_1 \la_1 \lst^2+\frac{1}{3} \la_2 \lst^4 \right),  \\
a_5 &=& \frac{1}{360} \left[192 a_1^3+66 a_1^2 \lst^2 \la_1+2 \lst^6 \la_1 \la_2+ a_1 \lst^4(3 \la_1^2+44 \la_2) \right]. \\
&& \ldots \nonumber 
\ean
At this stage, $a_1$, $\la_1$ and $\la_2$ are free parameters. Now we need to impose the Friedmann equations and see what mater content is required. Let us start with the first Friedmann equation \eqref{fried12}, working up to $O(t^5)$ for simplicity. We get
\be
\k^2 \rho (t) = \frac{1}{\lst^2} \left[ D_0+D_2 \left(\frac{t}{\lst} \right)^2+D_4 \left(\frac{t}{\lst} \right)^4 +O\left(\frac{t}{\lst} \right)^6 \right] ,\label{rhoseries}
\ee
with the coefficients
\ba
D_0 &=& -2 c_0 (|\la_1| \lst^2)^{\g-2}\left(3 a_1+2 \lst^2 \frac{\la_2}{\la_1} \right)\left[3 a_1(\g-1)+2(\g-3) \lst^2 \frac{\la_2}{\la_1} \right] \, , \\
D_2 &=& 3 a_1^2-6 a_1 c_0 (|\la_1| \lst^2)^{\g-2}(-2a_1+\lst^2 \la_1) \left(3 a_1+2\lst^2 \frac{\la_2}{\la_1} \right), \\
D_4 &=&  -\frac{a_1}{3} \left[3a_1 (4a_1+\lst^2 \la_1) +2 \lst^4 \la_2 \right]-\frac{c_0}{36} (\lst^2 |\la_1|)^{\g-2} \left\{\vphantom{\frac11}-58752a_1^5(\g-2) (\lst^2 \la_1)^{-1} \right. \nonumber \\
&& \left.+72 a_1^4 (1294-363 \g) +4 \lst^8 \la_2 (3 \la_1^2+98 \la_2)+3 a_1^2 \lst^4 \left[(510-9 \g) \la_1^2+4(1604-327 \g) \la_2 \right] \right.\nonumber \\
&& \left. -18 a_1^3 \lst^2 \left[(-1316+195 \g) \la_1+8(-250+121 \g) \frac{\la_2}{\la_1} \right] \right. \nonumber \\
&& \left. +6 a_1 \lst^6 \left[3 \la_1^3+(268-3 \g) \la_1 \la_2+16(24-11 \g)\frac{\la_2^2}{\la_1} \right]  \right\}.
\ea
On the other hand, from the second Friedmann equation \eqref{fried22} we find
\be
\k^2 (3p-\rho) = \frac{1}{\lst^2} \left[E_0+E_2 \left(\frac{t}{\lst} \right)^2+E_4 \left(\frac{t}{\lst} \right)^4 +O\left(\frac{t}{\lst} \right)^6 \right], \label{traceseries}
\ee
where
\ba
E_0 &=& -6 a_1 +4c_0 (\lst |\la_1|)^{\g-2} \left(3 a_1 +2 \lst^2 \frac{\la_2}{\la_1} \right) \left[6 a_1 (\g-2)+\lst^2 \left(3 \la_1+4 (\g-3) \frac{\la_2}{\la_1} \right) \right], \\
E_2 &=& \left(3 a_1+2 \lst^2 \frac{\la_2}{\la_1} \right) \left[l_*^2 \la_1-2c_0(\lst^2 |\la_1|)^{\g-2} \left(18 a_1 (\g-2) \lst^2 \la_1 +\lst^4 (3 \la_1^2+4(3\g-7)\la_2)\right) \right], \nonumber \\
\ea
and a longer expression for $E_4$.  

If we try to make some of the terms vanish exactly to simplify the matter content, in general one gets a behaviour similar to the one found in the previous sub-section. For instance, one can cancel the $E_2$ term for any $\g$ by fixing the constants to
\ben 
\frac{\la_2}{\la_1} = -\frac{3}{2}\frac{a_1}{\lst^2}\,,
\een
which also enforces $E_0=-6a_1$, $D_0 = 0$ and $D_2=3a_1^2$. Since it must be $\dot{H}>0$ at $t=0^+$, we have $a_1>0$. Then, near the bounce the equations of motion become
\ben
\k^2 \rho = 3 \frac{a_1}{\lst^2} \left(\frac{t}{\lst}\right)^2+O(t^4)\,,\qquad \k^2 (3p-\rho) = -6 \frac{a_1}{\lst^2}+O(t^4)\,,
\een
leading to a negative barotropic index $w\ll-1$ that diverges at the bounce:
\ben
w=-\frac{2}{3}\left(\frac{\lst}{t}\right)^2+\frac13+O(t^4)\,.
\een
This method of cancelling some coefficients does not work well with fractional operators. As we have already mentioned, in \cite{Biswas:2010zk} specific form factors were taken such that the 00 component of the modified Einstein equations and the trace equation only require, respectively, radiation plus a cosmological constant and a cosmological constant. This is something we cannot achieve for \eqref{gravac} and we cannot simply study the sign of just one radiation contribution.

Substantial progress is made when we consider a global cosmological fluid and see whether we can have a bounce satisfying the NEC and the WEC, \Eqqs{nec} and \Eq{wec}. As briefly recalled in appendix~\ref{appD}, both conditions are violated in a bounce in ordinary Einstein gravity and the WEC is also violated in Starobinsky $R+R^2$ gravity, on a flat FLRW background. Therefore, if we could find a non-trivial corner in the parameter space of FQG such that both conditions are satisfied, then we would obtain strong evidence that the nonlocal fractional operators can sustain a bounce with ordinary matter.

\begin{figure}[t]
    \centering
    \includegraphics[width=\linewidth]{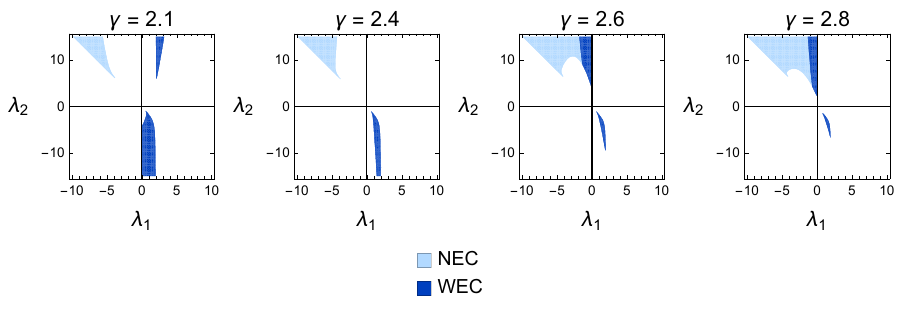}
    \caption{Regions in the $(\la_1, \la_2)$-plane where the WEC (which implies the NEC) is satisfied (dark blue areas) and where only the NEC is verified (light blue areas).}
    \label{fig3}
\end{figure}

\begin{figure}
    \centering
    \includegraphics[width=0.7\linewidth]{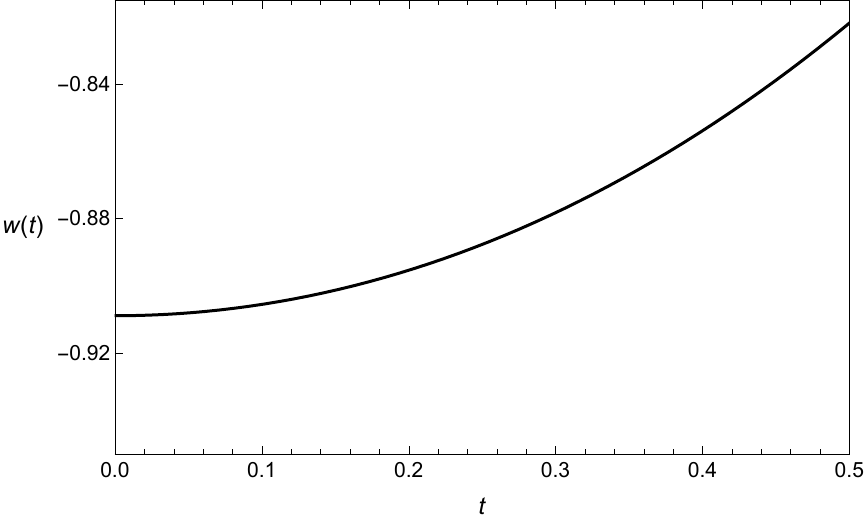}
    \caption{Equation of state obeying the WEC \Eq{wec} in the example $(\g, \la_1, \la_2) = (2.1, 1, -10)$.}
    \label{fig4}
\end{figure}

We proceed as follows. First, we consider only the $O(t^0)$ and $O(t^2)$ terms in \eqref{rhoseries} and \eqref{traceseries} around the bounce time $t = 0$. We work in $\lst=1$ units and fix $c_0 = 1=a_1$ without any loss of generality. Sufficient conditions to verify the NEC \Eq{nec} at $t \ll 1$ are
\be
4D_0+E_0 \geq 0 \, , \qquad 4D_2+E_2 \geq 0 \, , 
\ee
while the WEC is satisfied if, in addition,
\be
D_0 \geq 0 \, , \qquad D_2 \geq 0 \, .
\ee
In figure \ref{fig3}, we show the regions in the $(\la_1, \la_2)$-plane where these conditions are met.
These regions have zero extension in Einstein ($NEC$) and Starobinsky gravity ($WEC$). As an example of a WEC-preserving fractional scenario, we can take $(\g,\la_1,\la_2) = (2.1,1,-10)$. Inserting these numbers in \eqref{Hseries}, one obtains
\be
H(t) = t+\frac{5}{18}t^3+O(t^5)\,,
\ee
which corresponds to the scale factor
\be
\frac{a(t)}{a(0)} = 1+\frac{t^2}{2}+O(t^4)  \, .
\ee
The equation of state is plotted in figure \ref{fig4} and, indeed, the barotropic index $w$ does not drop below $-1$.


\section{de Sitter solution} \label{dynsys}

The exponential de Sitter profile $a(t) = a_0\,\exp(H t)$ with $H = {\rm const}$ is an exact solution of \eqref{gen00eq} and \eqref{gentreq} since all the modifications to the Einstein equations cancel as $R_{\mu \nu} = 3H^2 g_{\mu \nu}$ is covariantly constant (we are considering a metric-compatible connection). As usual, we need to sustain this accelerated expansion with a cosmological constant $\Lambda = 3H^2$. This is a solution of the scalar \emph{Ansatz} \eqref{ansatzR} where
\be\label{lala}
\frac{\la_2}{\la_1} = -\Lambda\,,
\ee
so that $\la_2$ and $\la_1$ must have opposite signs. 

To study the stability of this solution, we transform the cosmological equations into a first-order system of differential equations similarly to what done in \cite{Biswas:2010zk}. We use the definitions
\ba 
\epsilon \coloneqq -\frac{\dot{H}}{H^2}, \quad \eta \coloneqq \frac{\dot{\epsilon}}{H}, \quad \Sigma \coloneqq \frac{\la_1}{H^2}, \quad \Omega_{\rm r} \coloneqq \frac{\k }{3H^2} \rho_{\rm r}, \quad \Lambda = \k^2\rho_\Lambda \,,
\ea
where we add a radiation component for completeness. Note that $\epsilon$ is the usual slow-roll parameter but $\eta$ is not (the ordinary second slow-roll parameter $-\ddot\phi/(H\dot\phi)$ is not suitable in pure de Sitter) and that all parameters but $\Lambda$ are dimensionless. It is easy to check that
\be
\dot{H} = -\epsilon H^2 \, ,  \qquad
\ddot{H} = H^3 (2 \epsilon^2-\eta) \, , \qquad
\dddot{H} = H^4 (-6 \epsilon^3+7 \epsilon \eta)-H^3 \dot{\eta} \, .
\ee
From the first Friedmann equation \eqref{fried12} (Balakrishnan--Komatsu representation), we find an expression for $\eta' \coloneqq \rmd\eta/\rmd \log{a}= \dot{\eta}/H$:
\ba
\eta' = -6 \epsilon^3+7 \epsilon \eta +\frac{f(\epsilon, \eta, \Sigma)+ \left(-1+\frac{\Lambda}{3 \la_1} \Sigma+\Omega_r \right) \frac{\Sigma^2}{2 c_0 (\la_1 \lst^2)(|\la_1| \lst^2)^{\g-2}}}{2 \Sigma +3(\g-2)(\epsilon-2)} \label{dynsys1}\, ,
\ea
where
\ba 
f(\epsilon,\eta,\Sigma) &=& 3(\g-2)\left[2 \epsilon^3+4 \epsilon^4-4 \epsilon^2(8+\eta)+\eta(14+\eta)+\epsilon(24+\eta)  \right]\nn
&&+ \left[3 \epsilon(-2+9 \epsilon)-8 \eta  \right] \Sigma-2(-2+\epsilon) \Sigma^2 \nonumber \\
&& +\frac{2 \la_2 \left[-3 \g(-2+\epsilon)+2(-9+3\epsilon+\Sigma) \right] }{3 \la_1^2} \Sigma^2 + \frac{4 (\g-3)\la_2^2}{3 \la_1^4} \Sigma^3 \, .
\ea
From \eqref{ansatzR}, one obtains another first-order equation for $\epsilon$:
\ba 
\epsilon'\! &=&\! \frac{1}{7}\! \left[18 \epsilon^2 -12 \epsilon +(2-\epsilon) \Sigma +\frac{2}{3} \frac{\la_2}{\la_1^2} \Sigma^2  -\frac{f(\epsilon, \eta, \Sigma)+ \left(-1+\frac{\Lambda}{3 \la_1} \Sigma+\Omega_r \right) \frac{\Sigma^2}{2 c_0 (\la_1 \lst^2)(|\la_1| \lst^2)^{\g-2}}}{2 \Sigma +3(\g-2)(\epsilon-2)} \right]\!. \nonumber \label{dynsys2}\ \\
&&
\ea
From the definition of $\Sigma$ and from energy conservation of radiation, we can close the system $(\eta',\epsilon',\Sigma',\Omega'_r)^T = F(\eta,\epsilon,\Sigma,\Omega_r)^T$ with
\ba
\Sigma'&=& 2\epsilon \Sigma \, , \label{dynsys3}\ \\
\Omega_{\rm r}' &=& 2 (\epsilon-2) \Omega_{\rm r} \, . \label{dynsys4}\
\ea
Equations \eqref{dynsys1} and \eqref{dynsys2}--\eqref{dynsys4} fully specify the dynamical system.

Let us show that a de Sitter universe is a fixed point of this system. We have $\epsilon = \eta = \Omega_{\rm r} = 0$ and $\Sigma = 3 \la_1/\Lambda$, $\Lambda = 3H^2 = \text{const}$, to which we add \Eq{lala}. Then, one can check that \eqref{dynsys1}--\eqref{dynsys4} vanish identically (note that $f(0,0,3\la_1/\Lambda) = 0$), so that de Sitter is a fixed point. 


To see its stability, for simplicity we work in $\lst = 1$ units. The Jacobian matrix at the de Sitter fixed point is given by
\ba
\mathcal{J}_{\text{dS}} &\coloneq& \left. \frac{\partial F}{\partial(\eta,\epsilon,\Sigma,\Omega_r) } \right|_{\eta=0,\epsilon=0,\Sigma=\frac{3\la_1}{\Lambda},\Omega_{\rm r} = 0} \nonumber \\
&=& 
\begin{pmatrix}
-4+\frac{3(\g-2)}{2-\g+\la_1/\Lambda} && -3-3 \frac{\la_1}{\Lambda}+\frac{9(\g-2)}{\la_1/\Lambda+2-\g} && \frac{|\la_1|^{2-\g}-8 c_0 \left[\la_1+(3-\g) \Lambda  \right]}{4 c_0 \left[\la_1+(2-\g) \Lambda \right]} && \frac{3 \la_1 |\la_1|^{2-\g}}{4c_0 \Lambda \left[\la_1+(2-\g) \Lambda \right]} \\
\frac{\g-2-(4/7)(\la_1/\Lambda)}{\g-2-\la_1/\Lambda} && \frac{9 \la_1}{7 \Lambda} \frac{1}{\g-2 -\la_1/\Lambda} && \frac{8c_0 \Lambda -|\la_1|^{2-\g}}{28 c_0 \left[(2-\g) \Lambda+\la_1 \right]} && \frac{3}{28 c_0} \frac{\la_1}{\Lambda} \frac{|\la_1|^{2-\g}}{\g-2-\la_1/\Lambda} \\
0 && \frac{6 \la_1}{\Lambda} && 0 && 0 \\
0 && 0 && 0 && -4 
\end{pmatrix}\!.
\nonumber \\
&& 
\label{dSmatrix}
\ea

\begin{figure}[ht!]
\centering

\begin{minipage}{0.48\textwidth}
\centering
\includegraphics[width=\linewidth]{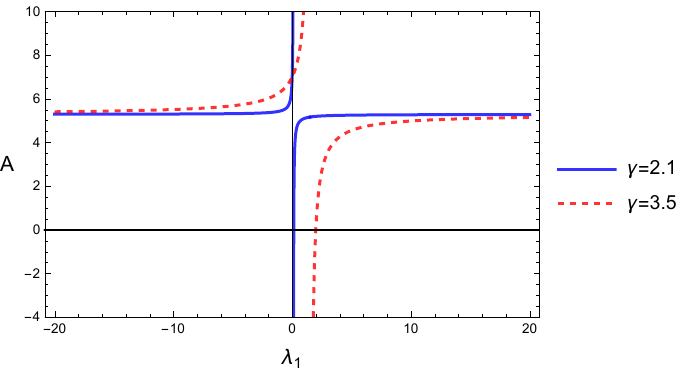}
\end{minipage}
\hfill
\begin{minipage}{0.48\textwidth}
\centering
\includegraphics[width=\linewidth]{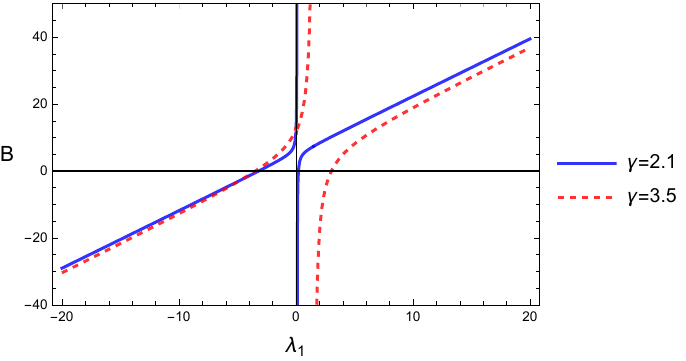}
\end{minipage}

\vspace{0.3cm}

\begin{minipage}{0.48\textwidth}
\centering
\includegraphics[width=\linewidth]{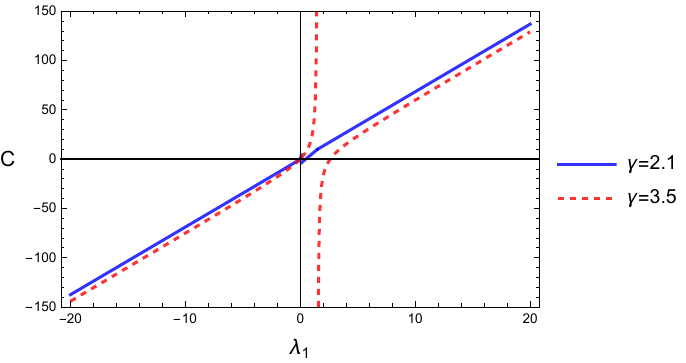}

\end{minipage}
\hfill
\begin{minipage}{0.48\textwidth}
\centering
\includegraphics[width=\linewidth]{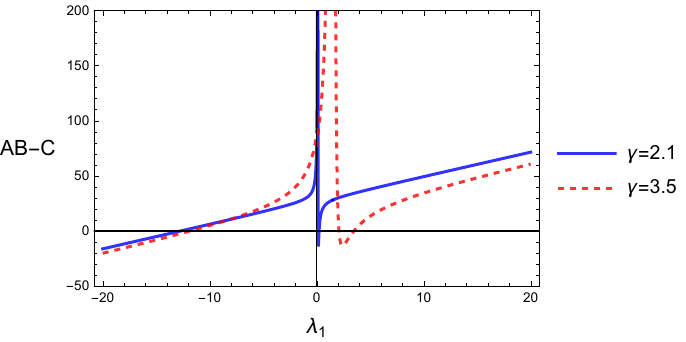}
\end{minipage}

\caption{Coefficients of the Routh--Hurwitz criterion as functions of $\lambda_1$ for two values of $\gamma$.\label{fig5}}
\end{figure}

We now focus on whether de Sitter is a stable fixed point (i.e., when all the corresponding eigenvalues of \eqref{dSmatrix} have negative real part). The characteristic polynomial is 
\be
p(\mu)
=
(\mu+4)
\left(
\mu^3+A \mu^2+B \mu+C \right) \, ,
\ee
where ($\mathcal{J}_{ij}$ are the $i,j$ entries in \eqref{dSmatrix})
\ba
A &=& -\left(\mathcal{J}_{11}+\mathcal{J}_{22}\right) \, , \\
B &=& \mathcal{J}_{11}\mathcal{J}_{22}
-\mathcal{J}_{12}\mathcal{J}_{21}
-\mathcal{J}_{32}\mathcal{J}_{23} \, , \\
C &=& \mathcal{J}_{32}\left(\mathcal{J}_{11}\mathcal{J}_{23}-\mathcal{J}_{13}\mathcal{J}_{21} \right) \, .
\ea
Clearly, $\mu=-4$ is an eigenvalue regardless of the other parameters. Focusing on the cubic polynomial in brackets, one can use the {Routh--Hurwitz criterion} to check whether all eigenvalues have negative real part. One needs to satisfy the following four conditions:
\be 
A > 0 \, , \quad  B > 0 \, , \quad C > 0 \, , \quad  AB -C>0  \, . \label{stabcond}
\ee
Let us work in $\Lambda=1$ units. In that case, we obtains
\ba
A &=& \frac{37}{7}+\frac{12(\g-2)}{7(-2+\g-\la_1)}\,, \label{Acoef}\\
B &=& \frac{3 \left\{2 c_0 \left[56+26 \la_1+4 \la_1^2-7 \g (4+\la_1) \right]+\la_1^3 |\la_1|^{-\g} \right\}}{14 c_0 (2-\g+\la_1)}\,,  \label{Bcoef}\\
C &=& \frac{12 \la_1 (14-7 \g +4 \la_1)}{7 (2-\g+\la_1)}\,. \label{Ccoef}
\ea
We start considering the case $c_0=1$. In figure~\ref{fig5}, we plot these parameters for two values of $\g$ in the range $2 < \g < 4$, as a function of $\la_1$. 

\begin{figure}[ht]
\begin{minipage}{0.48\textwidth}
\centering
\includegraphics[width=\linewidth]{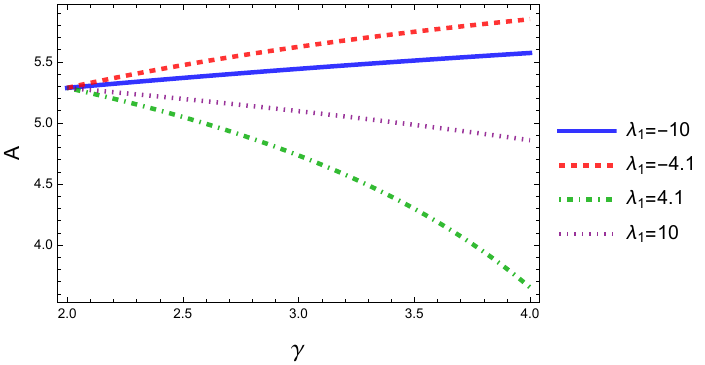}
\end{minipage}
\hfill
\begin{minipage}{0.48\textwidth}
\centering
\includegraphics[width=\linewidth]{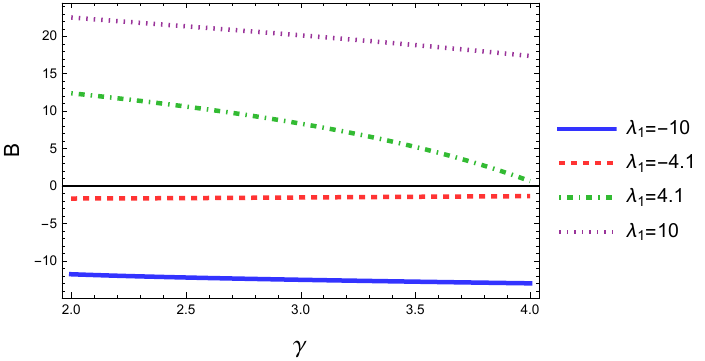}
\end{minipage}

\vspace{0.3cm}

\begin{minipage}{0.48\textwidth}
\centering
\includegraphics[width=\linewidth]{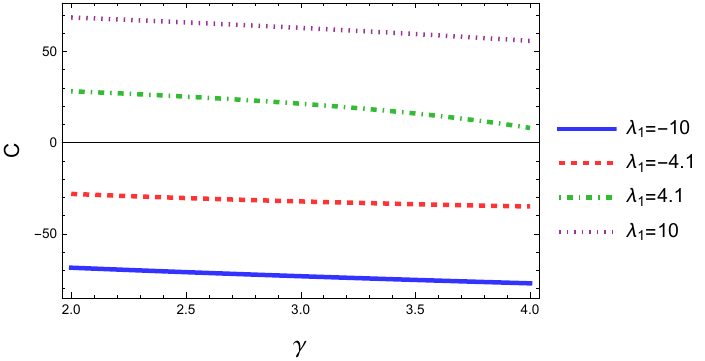}

\end{minipage}
\hfill
\begin{minipage}{0.48\textwidth}
\centering
\includegraphics[width=\linewidth]{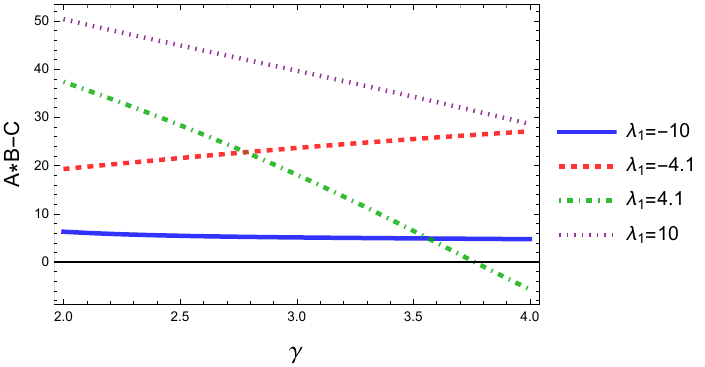}
\end{minipage}

\caption{
Coefficients of the Routh--Hurwitz criterion as functions of $\g$ for four values of $\lambda_1$. We have chosen $\la_1 = \pm 4.1$ because, when exploring the range of $2<\g<4$, one eventually hits the singular value $\la_1 = \g-2$ for $|\la_1| \leq 4$ in the coefficients \eqref{Acoef}--\eqref{Ccoef}.\label{fig6}
}
\end{figure}

The results are similar for different values of $\g$ and it is clear that the stability region (positive vertical axis) corresponds to $\la_1 > 0$, hence where the Balakrishnan--Komatsu and Fresnel representations coincide. Note the discontinuity at $\la_1 = \g-2$. In figure~\ref{fig6}, we study the $\g$-dependence for different values of $\la_1$. If $\la_1\gtrsim 4$, then all coefficients are positive in the whole $\g$ range chosen here; this condition would need to be enforced only if $\g$ took values close to 4.

We have considered the stability analysis in the case $c_0 = -1$ (which changes $B$) obtaining similar results. We conclude that the stability region corresponds to $\la_1>0$ in our range of $\g$.


\section{Discussion and conclusions} \label{disc}


\subsection{Summary of the main results}

In this work, we have begun to explore the cosmology of fractional gravity \eqref{gravac} for the first time. The high-energy behaviour of this theory, which has recently been introduced after a long propaedeutic study of multi-fractional spacetimes, is still under investigation but its renormalizability and unitarity properties \cite{Calcagni:2022shb,Calcagni:2026sfx} make it a promising candidate for quantum gravity. Among the novel results we obtained, we have written down the covariant equations of motion for self-adjoint fractional form factors and the Friedmann equations both in their general form and for the \emph{Ansatz} \Eq{ansatzR} in the presence of only an $R\cF_R(\B)R$ term in the action ($c_2=0$). 

To extract some physics through cosmological solutions, we have studied an hyperbolic cosine bounce, which is supported by unconventional energy components such as phantom ($w<-1$, violating both the NEC and the WEC) or ghost ($\rho<0$, violating the WEC) fluids, depending on the sign of $c_0$ (respectively, positive and negative). In the latter case, the universe passes through an instant of zero energy after which the total energy density is positive. The analysis of a more general bounce expressed as a temporal series shows that there do exist matter configurations respecting both the null and the weak energy conditions \Eq{nec}--\Eq{wec} for certain values of the fractional exponent $\g$. Although the parameter space of the theory is restricted by these constraints, it is not fine-tuned and bounces with ordinary energy-momentum components can occur naturally. All of this was found under the \emph{Ansatz} \Eq{ansatzR}, which clearly has more limitations than in NLQG-inspired models with entire form factors \cite{Biswas:2005qr,Biswas:2010zk,Koshelev:2014voa,Koshelev:2016xqb,Koshelev:2012qn,Biswas:2012bp}. Indeed, it is quite possible that fractional gravity admit other bounce solutions not obeying \eqref{ansatzR}. 

Moreover, we checked that de Sitter is an exact, stable solution, thus opening up the opportunity to realize accelerating cosmologies in FQG. Higher-curvature terms in the action typically sustain acceleration without the need of an inflaton, but a quantitative result was necessary to support expectations.


\subsection{Universality classes}

The above solutions are valuable also because they suggest the equivalence between the Balakrishnan--Komatsu representation \Eq{fracdalem2} and the Fresnel representation \Eq{fracdalem4} in the sense of belonging to the same universality class of operators. Different definitions of positive elliptic operators such as the fractional Laplacian are exactly equivalent \cite{Kwa17}. Much less is known about non-positive operators such as the fractional d'Alembertian but there are arguments pointing out that the Balakrishnan--Komatsu and the Fresnel representations, albeit mathematically inequivalent, should give the same physics at least in asymptotic regimes \cite{Calcagni:2025wnn,Briscese:2026jcf}. 

To understand this, let us recall that fractional operators can be composed in infinitely many ways that maintain the same asymptotic scaling in momentum space. For example, the operators of the family $\cK_\a(\B)=\B+\lst^{2(\g-1)}[(-\lst^2\B)^\a]^{\g/\a}$ parametrized by $\a\in\mathbb{R}$ all have the same scalings $\sim k^2$ in the IR ($\lst k\ll 1$) and $\sim (k^2)^\g$ in the UV ($\lst k\gg 1$) but can differ in the transient regime in-between according to the value of $\a$. Moreover, such value may introduce freedom in the representation of $\B^\a$, for instance when $\a=2$ as in the self-adjoint version of the theory, thus adding a second level of ambiguity. Despite these differences, the universality of the scaling has important consequences. All the operators $\cK_\a$ have exactly the same renormalizability properties for exactly the same range of $\g$ and, even if they may differ in their mathematical spectrum (in particular, they may have different complex poles), the physical spectrum can always be restricted to be the same \cite{Calcagni:2026sfx}. At the level of classical solutions of the dynamics, the asymptotic equivalence and the inequivalence in transient regimes are mirrored by a corresponding identical asymptotic behaviour at early and late times, with deviations expected at intermediate times. Complex systems with many characteristic scales are known to be adequately described in terms of these scales through the asymptotic behaviour of correlation functions. Examples are the renormalization group and effective field theory near fixed points \cite{Wilson:1971bg,Fisher:1974zz,Goldenfeld:1992qy}, turbulent, percolating and critical phenomena \cite{Frisch:1995zz,Barenblatt:1996bg}, chaotic systems \cite{Sornette:1997pb} and, closer to the subject of this paper, the wider set of multi-fractional classical and quantum field theories \cite{Calcagni:2016azd,Calcagni:2012rm,Calcagni:2016xtk,Calcagni:2017via}. 

Here we have compared the dynamics of FQG in the Balakrishnan--Komatsu and Fresnel representations and found that they not only lie in the same universality class but they actually give the same stable de Sitter and hyperbolic cosine bouncing solutions. This is a first example of an explicit comparison of two different representations. Whether the full equivalence between the two on the $\sech$ and de Sitter solutions is a peculiar coincidence or is a more general result for \emph{physical} stable solutions remains to be seen. 


\subsection{Comparison with multi-fractional models}

The classical dynamics of FQG is difficult to solve because it is nonlocal. A potential foothold in the problem can be gained applying the above concept of universality classes to the whole theory. In fact, FQG is only one among the various models falling into the multi-fractional paradigm where kinetic terms and spacetime measures exhibit anomalous scalings. We can compare our results with what is known on the early universe in these multi-fractional models. There are five classical models of this type in total, dubbed $T_1$, $T_v$, $T_q$, $T[\p^\g]$ and $T[\B^\g]$ \cite{Calcagni:2021ipd,Calcagni:2016azd}. The model $T[\B^\g]$ is FQG and is what we studied here. Within each model, the measure weight and the kinetic term may vary depending on whether one considers a UV purely fractional geometry (e.g., kinetic term $\sim\B^\g$) or one with both UV and IR limits in place (e.g., kinetic term $\sim\B+\B^\g$).\footnote{To differentiate between models with purely fractional and multi-scale kinetic terms, the notations $T[\p^\g]$, $T[\p+\p^\g]$, $T[\B^\g]$ and $T[\B+\B^\g]$ were employed in \cite{Calcagni:2021ipd}. We do not make this distinction here.} The classical models $T_1$, $T_v$ and $T_q$ are much simpler than $T[\B^\g]$ because they are local and one could use them as a shortcut guidance to probe the physics of $T[\B^\g]$ more easily.

Most of the research on this topic was conducted on the late-time acceleration of the universe \cite{Karami:2012ra,Lemets:2012fp,Shchigolev:2013jq,Chattopadhyay:2013mwa,Shchigolev:2015rei,Rami:2015kha,Maity:2016dfv,Debnath:2019isy,Jawad:2016piq,Sadri:2018lzz,Das:2018bxc,Maity:2019qbv,Ghaffari:2019qcv,Jawad:2019xdv,Calcagni:2020ads,Mamon:2020ocb,Barrientos:2020kfp,Landim:2021www,Shchigolev:2021lbm,Garcia-Aspeitia:2022uxz,Micolta-Riascos:2023mqo,Junior:2023fwb,deOliveiraCosta:2023srx,Marroquin:2024ddg,Jalalzadeh:2024qej,Micolta-Riascos:2024vbm,Kotal:2025hus} but we also have some knowledge on bouncing solutions and early acceleration \cite{Calcagni:2010bj,Shchigolev:2010vh,Calcagni:2013yqa,Calcagni:2016ofu,Jalalzadeh:2022uhl,Rasouli:2022bug,Rasouli:2024crg,Rasouli:2025qix,Oliveira:2026hjx,Rasouli:2026hfy,Calcagni:2016azd}:
\begin{itemize}
\item In the model $T_1$ with a non-trivial measure $\rmd^D x\,v(x)=\rmd^Dx\prod_\mu v_\mu(x^\mu)$ and ordinary derivatives $\p_\mu$, bouncing and non-bouncing solutions with a negative energy density (in the absence of a kinetic term $\p_\mu v\p^\mu v$ for the measure scalar weight \cite{Calcagni:2010bj,Rasouli:2025qix}) or violating the usual energy conditions (in the presence of such term \cite{Calcagni:2010bj}) are common. An inspection of the Friedmann equations shows that one can get acceleration from geometry without slow-rolling fields \cite{Calcagni:2016azd}. A quasi-de Sitter phase and the mechanism of inflation are generally possible \cite{Shchigolev:2010vh,Rasouli:2024crg,Oliveira:2026hjx,Rasouli:2026hfy}, even with a standard slow-rolling scalar field \cite{Rasouli:2024crg,Oliveira:2026hjx}.
\item In the model $T_v$ with a non-trivial measure and weighted derivatives $v^{-\b}\p_\mu(v^\b\,\cdot\,)$ (where $\b$ depends on the rank of the tensor on which this operator acts), de Sitter with a cosmological constant is not an exact solution but there is a vacuum bouncing solution that tends to de Sitter at late times \cite{Calcagni:2013yqa}. Again, the Friedmann equations indicate the possibility of acceleration sustained only by curvature \cite{Calcagni:2016azd} and the possibility of standard inflation in $T_v$ is unclear. 
\item In the model $T_q$ with non-trivial measure and $q$-derivatives $v_\mu^{-1}\p_\mu$, a bouncing solution tending to de Sitter exists in the presence of a cosmological constant \cite{Calcagni:2013yqa}. Curvature alone cannot sustain acceleration and solve the hot-big-bang problems in this model \cite{Calcagni:2016azd}. On the other hand, it was shown that inflation can take place with a conventional scalar-field mechanism but a non-trivial imprint in the CMB \cite{Calcagni:2013yqa,Calcagni:2016ofu} and that the big bang can be avoided if one takes into account the logarithmic oscillations in the measure at ultra-short scales \cite{Calcagni:2013yqa}. This structure is typical of the self-similar geometry of deterministic fractals and it could also appear in FQG after a modification of the kinetic term at scales below $\lst$ \cite{Briscese:2026jcf}:
\be\label{Klog}
\cF_2(\B)\to \cK_2(\B)\coloneqq\cF_2(\B)\sum_{n=0}^N\cos[n\Omega\ln(\lst^4\B^2)+\theta_n]\,,
\ee
where $\theta_n$ are phases and $\Om$ is a frequency parameter. We have not considered this version of the theory here and, therefore, we cannot comment on this further.
\item The theory $T[\p^\g]$ with ordinary measure and fractional derivatives has been the least studied due to the technical challenge these coordinate-dependent integro-differential operators $\p^\g$ pose \cite{Calcagni:2018dhp}. In some toy models, it was found that the de Sitter metric is a natural background in cosmology with fractional derivatives $\p_\mu^\a$ \cite{Jalalzadeh:2022uhl} but, contrary to standard cosmology, slow-roll inflation is realized in a background with power-law scale factor rather than in an exponentially accelerating one \cite{Rasouli:2022bug}.
\end{itemize}
To the best of our current understanding of the cosmology of multi-fractional models, $T_1$ seems to be the closest analogue of our findings in classical fractional gravity, where there are no bouncing vacuum solutions and the matter content can be highly exotic. The evidence gathered here of some correspondence $T_1\leftrightarrow T[\B^\g]$ at the classical cosmological level can stimulate a renewed and keener interest in the purely phenomenological scenarios based on $T_1$, which can serve as forerunners redirecting research on $T[\B^\g]$. For example, if an alternative to inflation in $T_1$ were to pass consistency and observational tests, this would indicate the opportunity to undertake a similar model building and analysis in $T[\B^\g]$.

It remains to be seen whether one can go beyond a superficial analogy and use the model $T_1$ as a genuine dual, perhaps in the exact holographic sense suggested in \cite[section~5]{Calcagni:2021ljs}. This would require to generalize the Caffarelli--Silvestre extension theorem \cite{Caffarelli:2007eci,Frassino:2019yip} valid for a non-hermitian kinetic term $(-\B)^\g$ to a self-adjoint, multi-scale kinetic operator $\sim\B+(\B^2)^{\g/2}$.


\subsection{Future research in fractional cosmology}

We conclude with three more invitations to future research on the cosmology with fractional operators. First and foremost, the current analysis has been conducted on the equations of motion coming from the action \Eq{gravac} but this is not quite the classical limit of the theory. In fact, the propagator contains complex-conjugate modes \cite{Calcagni:2026sfx} that include ghosts and must be projected out of the quantum spectrum of asymptotic states with the Anselmi--Piva prescription on scattering amplitudes \cite{Anselmi:2017yux,Anselmi:2017lia,Anselmi:2018tmf,Anselmi:2018bra,Anselmi:2019rxg,Anselmi:2021hab,Anselmi:2022toe,Anselmi:2025uzj,Anselmi:2025uda}. This prescription preserves the optical theorem and Lorentz invariance \cite{Anselmi:2025uzj} and, although originally devised to handle problematic poles, it can be adapted to the continuum of modes of fractional QFT \cite{Calcagni:2026sfx,Anselmi:2026hvr}. Complex-conjugate modes are made purely virtual and integrated out of the dynamics, so that they disappear from the actual classical equations of motion in exchange for nonlocal interactions \cite{Anselmi:2018tmf,Anselmi:2018bra,Anselmi:2019rxg,Anselmi:2025uda}. We have not computed these classicized equations of motion but this would be the next important step to understand whether the bouncing and accelerating solutions found here survive or are modified.

Second, instead of trying closed forms for the bouncing scale factor $a(t)$, one could employ series expansions as in section~\ref{genbou} but for the stronger \emph{Ansatz} \eqref{ansatzricci}, which would allow one to consider the full dynamics with $c_2\neq 0$. In alternative, the diffusion method \cite{Calcagni:2025wnn} allows one to get access to a much wider portion of the space of solutions.

Third, since de Sitter is a solution of the theory, one could study cosmological perturbations during an inflationary phase. This problem is not easy in nonlocal theories but there are some results in the case of entire form factors \cite{Koshelev:2016xqb}. These future analyses should be carried out with the classicized equations of motion.


\section*{Acknowledgments}

The authors thank A.~De Felice and S.~Tsujikawa for useful discussions at the beginning of this project. I.S.G.\ is supported by the Spanish Ministry of Science, Innovation and Universities through FPU grant FPU24/02476. G.C.\ is supported by grant PID2023-149018NB-C41 funded by the Spanish Ministry of Science, Innovation and Universities MCIN/AEI/10.13039/ 501100011\-033. His work was made possible also through the support of the WOST, \href{https://withoutspacetime.org}{WithOut SpaceTime project}, supported by Grant ID 63683 from the John Templeton Foundation. The opinions expressed in this work are those of the authors and do not necessarily reflect the views of the John Templeton Foundation.


\appendix
\addtocontents{toc}{\protect\setcounter{tocdepth}{1}}


\section{Derivation of the equations of motion} \label{appA}

In this section, we present some details on the derivation of the modified Einstein equations. In section~\ref{appA1}, we consider a general quadratic action with arbitrary form factors \cite{Calcagni:2021aap}, while in sections~\ref{appA2}--\ref{appA3} we particularize for the two representations of the fractional d'Alembertian explored in the main text.


\subsection{Variations of a general quadratic action} \label{appA1}

For a general quadratic action with arbitrary form factors,
\ba
S &=& \frac{1}{2 \, \k^2} \int d^4 x \sqrt{|g|} \left[R+R \cF_R(\B) R+ G_{\mu \nu} \mathcal{F}_2 (\B)R^{\mu \nu} \right]+S_{\rm m}\nn
&\eqqcolon& \frac{1}{2 \, \k^2} \int d^4 x \sqrt{|g|} \mathcal{L}_g+S_{\rm m} \, ,
\ea
one can compute the modified Einstein equations of motion as usual:
\ba
\k^2 T_{\mu \nu} &=& \frac{2 \k^2}{\sqrt{|g|}} \frac{\de \left(\sqrt{|g|} \mathcal{L}_g \right)}{\de g^{\mu \nu}} \nonumber \\ 
&=& \frac{1}{\sqrt{|g|}} \left\{\frac{\de \left(\sqrt{|g|} R \right)}{\de g^{\mu \nu}}+\frac{\de \left[\sqrt{|g|} R \cF_R(\B) R \right]}{\de g^{\mu \nu}}+\frac{\de \left[\sqrt{|g|} G_{\a \b} \cF_2(\B) R^{\a \b} \right]}{\de g^{\mu \nu}} \right\},\label{varact}
\ea
where the matter energy-momentum tensor is
\be 
T_{\mu \nu} \coloneqq -\frac{2}{\sqrt{|g|}} \frac{\de S_{\rm m}}{\de g^{\mu \nu}} \, . \label{emt}
\ee
The first term in \eqref{varact} yields the Einstein--Hilbert contribution $G_{\mu\nu}=R_{\mu\nu}-(1/2)g_{\mu\nu}R$. The second term is
\ba
\frac{1}{\sqrt{|g|}} \frac{\de \left[\sqrt{|g|} R \cF_R (\B) R \right]}{\de g^{\mu \nu}} &=& -\frac{1}{2}  g_{\mu \nu}R \cF_R(\B) R+2  \left(R_{\mu \nu}+g_{\mu \nu} \B-\N_\mu \N_\nu \right) \cF_R (\B) R \nonumber \\
&&+ R  \frac{\de \cF_R (\B)} {\de g^{\mu \nu}}R  \, ,  \label{varact1}
\ea
while the last one yields
\ba
\frac{1}{\sqrt{|g|}} \frac{\de \left[\sqrt{|g|}G_{\mu \nu} \mathcal{F}_2(\B)R^{\mu \nu} \right]}{\de g^{\mu \nu}}  &=& - \frac{1}{2}  g_{\mu \nu} G_{\a \b} \mathcal{F}_2(\B) R^{\a \b}+2 G^\s_\mu \mathcal{F}_2(\B)G_{\nu \s} \nonumber \\ 
&&+  \mathcal{F}_2(\B) \B G_{\mu \nu}+g_{\mu \nu} \N^\s \N^\t \mathcal{F}_2(\B)G_{\s \t}-2 \N^\s \N_\mu \mathcal{F}_2(\B) G_{\nu \s}  \nonumber \\
&&+\frac{1}{2}  \left[G_{\mu \nu} \mathcal{F}_2(\B)R+R\mathcal{F}_2(\B) G_{\mu \nu}  \right]+  G_{\s \t}   \frac{ \de \mathcal{F}_2(\B)}{\de g^{\mu \nu}} R^{\s \t} \, . \nonumber \\  \label{varact2}
\ea
In the case of \eqref{gravac}, we have $\mathcal{F}_i (\B) = \lst^2 (\lst^2 \dotBox)^{\g-2}$, so that we must compute the variation of \eqref{fracdalem2} or \eqref{fracdalem4}, depending on the representation we choose. 


\subsection{Self-adjoint Balakrishnan--Komatsu representation} \label{appA2}

Let us start with \eqref{fracdalem2} and set $\lst=1$. Its variation is
\ba
\de {\dotBox}^{\g-2} &=& \frac{1}{\Gamma(n-\g/2+1)} \int_0^{+ \infty} \rmd \t \, \t^{n-\frac{\g}{2}} \left[\de \left(\B^{2n}\right)\rme^{- \t \B^2}+\B^{2n} \de \left(e^{-\t \B^2} \right) \right] \nonumber \\
&=& \frac{1}{\Gamma(n-\g/2+1)} \int_0^{+ \infty} \rmd \t \, \t^{n-\frac{\g}{2}} \left\{\sum_{l=0}^{2n-1} \B^l (\de \B) \B^{2n-l-1}\rme^{-\t \B^2} \right. \nonumber \\
&& \, \left. -\B^{2n} \int_0^\t \rmd q \,  \rme^{-q \B^2} \left[(\de \B) \B+\B (\de \B) \right] \rme^{(q-\t)\B^2} \right\},
\ea
where we used Duhamel's identity
\be
\de \rme^{\t A} = \int_0^\t \rmd q \rme^{q A}(\de A)\rme^{(\t-q)A} \, ,
\ee
for the operator $A = -\B^2$. The last term in \eqref{varact1} is proportional to
\be
R \, (\de{\dotBox}^{\g-2}) \, R = \de g^{\mu \nu} \sum_{i = 1}^3 \vartheta_{\mu \nu}^{(i)} \left(A,B \right)+O(\N) \coloneqq \de g^{\mu \nu}\vartheta_{\mu \nu}(A,B) +O(\N)\, ,
\ee
where $O(\N)$ is a total derivative that does not contribute to the equations of motion and
\ba
\hspace{-1.5cm}&& \vartheta_{\mu \nu}^{(1)} \left(A,B \right) \coloneqq \frac{1}{\Gamma(n-\g/2+1)} \int_0^{+\infty} \rmd \t \,  \t^{n-\frac{\g}{2}} \sum_{l = 0}^{2n-1} \bar{\vartheta}_{\mu \nu}(\B^l A, \B^{2n-l-1} \rme^{- \t \B^2} B) \,,   \label{vartheta1}\\
\hspace{-1.5cm}&& \vartheta_{\mu \nu}^{(2)} \left(A,B \right)\coloneqq -\frac{1}{\Gamma(n-\g/2+1)} \int_0^{+\infty} \rmd \t \,  \t^{n-\frac{\g}{2}} \int_0^\t 
dq \, \bar{\vartheta}_{\mu \nu} \left(\B^{2n} \rme^{-q \B^2} A, \B \rme^{(q-\t) \B^2 } B \right), \label{vartheta2}\\
\hspace{-1.5cm}&& \vartheta_{\mu \nu}^{(3)} \left(A,B \right)\coloneqq -\frac{1}{\Gamma(n-\g/2+1)} \int_0^{+\infty} \rmd \t \,  \t^{n-\frac{\g}{2}} \int_0^\t \rmd q \,  \bar{\vartheta}_{\mu \nu} \left(\B^{2n+1} \rme^{-q \B^2} A,  \rme^{(q-\t) \B^2 } B \right), \label{vartheta3}
\ea
with the $A (\de \B) B$ variation \cite{Calcagni:2018lyd}
\be
\bar{\vartheta}_{\mu \nu}(A,B) \coloneqq -\N_{(\mu} B \N_{\nu)} A+ \frac{1}{2} g_{\mu \nu} \left(A \B B+ \N_\rho A \N^\rho B \right) \,.\label{barvartheta}
\ee
The last term in \eqref{varact2} is completely analogous:
\be
\hspace{-.2cm}R_{\s  \t} \, (\de{\dotBox}^{\g-2}) \, G^{\s \t} = \de g^{\mu \nu} \sum_{i = 1}^3 \Theta_{\mu \nu}^{(i)} \big(R_{\a \b},G^{\a \b} \big)+O(\N) \coloneqq \de g^{\mu \nu} \Theta_{\mu \nu} \big(R_{\a \b},G^{\a \b} \big)+O(\N)\,,
\ee
where
\ba
\Theta_{\mu \nu}^{(1)} \big(A_{\a \b},B^{\a \b} \big) &\coloneqq& \frac{1}{\Gamma(n-\g/2+1)} \int_0^{+\infty} \rmd \t \,  \t^{n-\frac{\g}{2}} \sum_{l = 0}^{2n-1} \bar{\Theta}_{\mu \nu}\left(\B^l A_{\a \b}, \B^{2n-l-1} \rme^{- \t \B^2} B^{\s \t} \right),  \nn\\
\Theta_{\mu \nu}^{(2)} \big(A_{\a \b},B^{\a \b} \big) &\coloneqq& -\frac{1}{\Gamma(n-\g/2+1)}\nn
&&\times\int_0^{+\infty} \rmd \t \,  \t^{n-\frac{\g}{2}} \int_0^\t 
\rmd q \, \bar{\Theta}_{\mu \nu} \left(\B^{2n} \rme^{-q \B^2} A_{\a \b}, \B \rme^{(q-\t) \B^2 } B^{\a \b} \right), \\
\Theta_{\mu \nu}^{(3)} \big(A_{\a \b},B^{\a \b} \big) &\coloneqq& -\frac{1}{\Gamma(n-\g/2+1)}\nn
&&\times\int_0^{+\infty} \rmd \t \,  \t^{n-\frac{\g}{2}} \int_0^\t
\rmd q \,  \bar{\Theta}_{\mu \nu} \left(\B^{2n+1} \rme^{-q \B^2} A_{\a \b},  \rme^{(q-\t) \B^2 } B^{\a \b} \right),
\ea
with the $A_{\a \b} (\de \B) B^{\a \b }$ variation \cite{Calcagni:2018lyd}
\ba
\bar{\Theta}_{\mu \nu}(A_{\a \b},B^{\a \b}) 
&\coloneqq& -\N_{(\mu} A_{\a \b} \N_{\nu)} B^{\a \b}+\frac{1}{4} g_{\mu \nu} \N_\rho \left(A_{\a \b} \N^\rho B^{\a \b}+B^{\a \b} \N^\rho A_{\a \b} \right) \nonumber \\
&& +\frac{1}{4} g_{\mu\nu} \N_\rho \left(A_{\a\b}\N^\rho B^{\a\b}-B^{\a\b}\N^\rho A_{\a\b}\right)+\N_\a \left(A_{(\mu\b}\N^\a B^\b_{\ \nu)}-B^\b_{( \nu}\N^\a A_{\mu )\b} \right)\nonumber\\
&& +\N_{\b}\left(B_{(\mu\a}\N_{\nu)} A^{\a\b}-A^{\a\b}\N_\nu B_{\mu\a}\right)+\N_\a \left(A_{(\mu\b}\N_{\nu)} B^{\b\a}-B^{\b\a}\N_{(\nu} A_{\mu)\b} \right) . \nonumber \\
\label{bartheta}
\ea


\subsection{Self-adjoint Fresnel representation} \label{appA3}

Proceeding as in the previous sub-section, we have
\be
R_{\s  \t} \, (\de\B_{\text{F}}^{\g-2}) \, G^{\s \t} = \de g^{\mu \nu} \sum_{i = 1}^2 \Xi_{\mu \nu}^{(i)} \big(R_{\a \b},G^{\a \b} \big) \coloneqq \de g^{\mu \nu} \Xi_{\mu \nu} \big(R_{\a \b},G^{\a \b} \big) \, ,
\ee
where
\ba
\Xi_{\mu \nu}^{(1)}(A_{\a \b},B^{\a \b} ) &\coloneqq& \frac{1}{\Gamma(n+2-\g) \cos{\frac{\pi(n+2-\g)}{2}}}\nn
&&\times \Re\int_0^\infty \rmd \t \t^{n+1-\g} \sum_{l=0}^{n-1} \bar{\Theta}_{\mu \nu} \left(\B^l A_{\s \t}, \B^{n-l-1} \rme^{\rmi \t \B} B^{\s \t}  \right),\\
\Xi_{\mu \nu}^{(2)}(A_{\a \b},B^{\a \b} ) &\coloneqq& \frac{1}{\Gamma(n+2-\g) \cos{\frac{\pi(n+2-\g)}{2}}}\nn
&&\times \Re \int_0^\infty \rmd \t \t^{n+1-\g} \,\rmi \int_{0}^\t \rmd q\, \bar{\Theta}_{\mu \nu} \left(\B^n \rme^{\rmi q \B} A_{\s \t}, \rme^{\rmi(\t-q) \B} B^{\s \t}  \right).
\ea
Similarly,
\be
R \, (\de{\dotBox}^{\g-2}) \, R = \de g^{\mu \nu} \sum_{i = 1}^3 \xi_{\mu \nu}^{(i)} \left(A,B \right) \coloneqq \de g^{\mu \nu}\xi_{\mu \nu}(A,B) \, ,
\ee
where
\ba
\xi_{\mu \nu}^{(1)}(A,B) &\coloneqq& \frac{1}{\Gamma(n+2-\g) \cos{\frac{\pi(n+2-\g)}{2}}}\, \Re \int_0^\infty \rmd \t \t^{n+1-\g} \sum_{l=0}^{n-1} \bar{\vartheta}_{\mu \nu} \left(\B^l A, \B^{n-l-1} \rme^{\rmi \t \B} B  \right),\nn\\
\xi_{\mu \nu}^{(2)}(A,B) &\coloneqq& \frac{1}{\Gamma(n+2-\g) \cos{\frac{\pi(n+2-\g)}{2}}}\, \Re \int_0^\infty \rmd \t \t^{n+1-\g}\, \rmi \int_{0}^\t \rmd q \bar{\vartheta}_{\mu \nu} \left(\B^n \rme^{\rmi q \B} A,\rme^{\rmi(\t-q) \B} B  \right).\nn
\ea


\section{Calculation of the Friedmann equations} \label{appB}

In this appendix, we provide some details of the computation of the Friedmann equations in the case of fractional gravity. We keep the spacetime dimension $D$ generic and use the notation of \cite{Gurses:2020kpv,Gurses:2024tka} to write the curvature tensors:
\be
R_{\mu\nu} \coloneqq P g_{\mu\nu}+Q u_\mu u_\nu \,,\qquad 
R = DP-Q\,.
\ee

We start by considering the first Friedmann equation obtained by taking the 00 component of \eqref{gen00eq}. The terms linear in curvature tensors yield
\ba 
&& \left(1+c_2 \lst^{2(\g-1)} {\dotBox}^{\g-2} \,  \B \right) G_{00} +2 c_0 \lst^{2(\g-1)} \left[g_{00} \B- \N_{0} \N_{0} \right] {\dotBox}^{\g-2} R  \nonumber \\
&& + \, c_2 \lst^{2(\g-1)} g_{00} \N^\s \N^\t {\dotBox}^{\g-2} G_{\s \t}-2 c_2 \lst^{2(\g-1)} \N^\s \N_{0} {\dotBox}^{\g-2} G_{0 \s} \nonumber \\
&&\qquad= 3 H^2+c_2 \lst^{2(\g-1)} {\dotBox}^{\g-2} \left[-\frac{\ddot{Q}}{2} + \frac{3}{2} H \dot{Q} +\left(1-\frac{D}{2} \right)(\ddot{P}-3H \dot{P})\right] \nonumber \\ 
&&\qquad\quad -6 \, c_0 \lst^{2(\g-1)} H \p_t {\dotBox}^{\g-2} (DP-Q) -c_2 \lst^{2(\g-1)} \N^\s \N^\t {\dotBox}^{\g-2} (Q u_\s u_\t) \nonumber \\ 
&&\qquad\quad +c_2 \, \lst^{2(\g-1)} \left(\frac{D}{2}-1 \right) \Box {\dotBox}^{\g-2} P-\frac{c_2}{2} \lst^{2(\g-1)} \B {\dotBox}^{\g-2} Q \nonumber \\ 
&&\qquad\quad -2 \, c_2 \lst^{2(\g-1)} \p_t^2 {\dotBox}^{\g-2} \left[P \left(1-\frac{D}{2} \right)+\frac{Q}{2}\right] -2 \, c_2 \lst^{2(\g-1)} \N^\s \N_0 {\dotBox}^{\g-2} (Q u_\s)  \nonumber \\ 
&&\qquad= 3 \left(H^2+\frac{\textsc{k}}{a^2} \right)+c_2 \lst^{2(\g-1)} {\dotBox}^{\g-2} \left[-\frac{\ddot{Q}}{2} + \frac{3}{2} H \dot{Q} +\left(1- \frac{D}{2}\right)(\ddot{P}-3H \dot{P})\right]  \nonumber \\ 
&&\qquad\quad -c_2 \lst^{2(\g-1)} \left[\N^\s \N^\t {\dotBox}^{\g-2} \left(Q u_\s u_\t \right)+2 \N^\s \N_0 {\dotBox}^{\g-2} \left(Q u_\s \right) \right] \nonumber \\ 
&&\qquad\quad + \lst^{2(\g-1)} \left[-6D c_0 H \p_t +c_2 \left(1-\frac{D}{2} \right) \left(\p_t^2+3H \p_t \right)\right] {\dotBox}^{\g-2} P \nonumber \\ 
&&\qquad\quad + \lst^{2(\g-1)} \left[6 c_0 H \p_t -\frac{c_2}{2} \left(\p_t^2+3H \p_t \right) \right] {\dotBox}^{\g-2} Q \,.
\ea 
The quadratic terms in curvature tensors in \eqref{gen00eq} give
\ba 
&& c_0 \lst^{2(\g-1)} \left(G_{0 0}+R_{0 0} \right) {\dotBox}^{\g-2}R-\frac{1}{2} c_2 \lst^{2(\g-1)} g_{0 0} G_{\s \t} {\dotBox}^{\g-2} R^{\s \t} \nonumber \\
&& +2c_2 \lst^{2(\g-1)} G_{0}^\s {\dotBox}^{\g-2} G_{0 \s}+ \frac{c_2}{2} \lst^{2(\g-1)} \left(G_{0 0} {\dotBox}^{\g-2}R+R {\dotBox}^{\g-2} G_{0 0} \right) \nonumber \\
&&\qquad= \lst^{2(\g-1)} \left[\frac{1}{2} \left(c_0-\frac{c_2}{2} \right) R {\dotBox}^{\g-2} R +\frac{c_2}{2} R_{ij} {\dotBox}^{\g-2}R^{ij}-\frac{3}{2} c_2 R_{00} {\dotBox}^{\g-2} R_{00} \right. \nonumber \\ 
&&\qquad\quad \left. + \left(2 \, c_0 -\frac{c_2}{2} \right)  \, R_{00} {\dotBox}^{\g-2}  R -\frac{c_2}{2} R {\dotBox}^{\g-2}  R_{00} \right] \nonumber \\
&&\qquad= \lst^{2(\g-1)} \left\{\frac{c_0}{2} D \left[(D-4)P+3Q \right]-\frac{c_2}{4}(D-4) \left[(D-2)P+Q \right] \right\} {\dotBox}^{\g-2} P \nonumber \\
&&\qquad\quad +\lst^{2(\g-1)} \left\{-\frac{c_0}{4}  \left[2(D-4)P+6Q \right]-\frac{c_2}{4} \left[(D-4)P+3Q \right] \right\} {\dotBox}^{\g-2} Q\,.
\ea
The final result is \eqref{fried1}, shown together with the second Friedmann equation \eqref{fried2} obtained by taking the trace.


\section{Derivation of \Eq{EinAns00}} \label{appC}

We focus on the self-adjoint Balakrishnan--Komatsu representation and the case $c_2 = 0$. Using the results in section~\ref{generalansatzsec}, the first term in \eqref{gen00eq} is
\ban 
&& 2 c_0 \lst^{2(\g-1)} \left[g_{\mu \nu} \B-\N_{(\mu} \N_{\nu)} \right] {\dotBox}^{\g-2} R \nonumber \\ 
&&\qquad= 2 \, c_0 \lst^2 (\lst^2 |\la_1|)^{\g-2} \left[g_{\mu \nu} \B-\N_{(\mu} \N_{\nu)} \right] \left(R+\frac{\la_2}{\la_1}D \right) \nonumber \\ 
&&\qquad= 2 \, c_0 \lst^2 (\lst^2 |\la_1|)^{\g-2} \left[\la_1 \left(R+\frac{\la_2}{\la_1} D \right)g_{\mu \nu}-\N_\mu \N_\nu R \right].
\ean
Similarly,
\ben 
c_0 \lst^{2(\g-1)} \left(G_{\mu \nu}+R_{\mu \nu} \right) {\dotBox}^{\g-2} R = c_0 \lst^2 (\lst^2 |\la_1|)^{\g-2}  \left(G_{\mu \nu}+R_{\mu \nu} \right) \left(R+\frac{\la_2}{\la_1}D \right).
\een
Now we simplify $\vartheta_{\mu \nu}$ in the range $2<\g<4$, so stat $n=1$ in \eqref{fracdalem2}. Starting from the sum in \eqref{vartheta1},
\ba 
&& \sum_{l=0}^{2n-1} \bar{\vartheta}_{\mu \nu} (\B^l R, \B^{2n-l-1} e^{-\t \B^2} R) \nonumber \\ 
&&\qquad= \bar{\vartheta}_{\mu \nu} \left(R, \B \rme^{-\t \B^2} R \right)+\bar{\vartheta}_{\mu \nu} \left(\B R,  \rme^{-\t \B^2} R \right) \nonumber \\ 
&&\qquad=  \bar{\vartheta}_{\mu \nu} \left[R, \la_1 \rme^{-\la_1^2 \t} \left(R+\frac{\la_2}{\la_1}D \right) \right]+\bar{\vartheta}_{\mu \nu} \left[\la_1 \left(R+\frac{\la_2}{\la_1}D\right), \rme^{- \la_1^2 \t}R +\frac{\la_2}{\la_1} (\rme^{-\la_1^2 \t}-1)D \right] \nonumber \\ 
&&\qquad=  2 \la_1 \rme^{-\la_1^2 \t} \bar{\vartheta}_{\mu \nu}(R,R)+ D \la_2 \bar{\vartheta}_{\mu \nu}(1,R) \rme^{-\la_1^2 \t} \nonumber \\ 
&&\qquad=  \la_1 \left[2 \bar{\vartheta}_{\mu \nu}(R,R)+ \frac{D}{2} \la_2 \left(R+\frac{\la_2}{\la_1}D \right) \, g_{\mu \nu} \right]\rme^{-\la_1^2 \t},  \label{ansthe1}
\ea
where we used \eqref{fR} in the second equality, the fact that \eqref{barvartheta} vanishes if the second argument is a constant and, in the last equality, that for the scalar \emph{Ansatz} \Eq{ansatzR} $\bar{\vartheta}(1,R) = ({\la_1}/{2})(R+{D\la_2}/{\la_1})g_{\mu \nu}$. Repeating the same for the integrals in $q$ in \eqref{vartheta2}--\eqref{vartheta3} with $n=1$, we obtain
\ba 
\hspace{-.9cm}\int_0^\t 
\rmd q \, \bar{\vartheta}_{\mu \nu} \left(\B^{2n} \rme^{-q \B^2} A, \B \rme^{(q-\t) \B^2 } B \right)\! &=&\!  \int_0^\t\rmd q \,  \bar{\vartheta}_{\mu \nu} \left(\B^{2n+1} e^{-q \B^2} R,  \rme^{(q-\t) \B^2 } R \right)\nonumber \\ 
&=&\! \la_1^3 \left[\bar{\vartheta}_{\mu \nu}(R,R)+\frac{D}{2}\la_2 \left(R+\frac{\la_2}{\la_1}D \right) g_{\mu \nu} \right] \t \rme^{-\la_1^2 \t}. \label{ansthe2}
\ea  
Using \eqref{ansthe1} and \eqref{ansthe2}, after performing the integrals in $\t$ one finally gets
\ba
\vartheta_{\mu \nu}(R,R) &=& \frac{|\la_1|^{\g-2}}{\la_1} \left\{2 \bar{\vartheta}_{\mu \nu}(R,R)+ \frac{D}{2} \la_2 \left(R+\frac{\la_2}{\la_1}D \right) \, g_{\mu \nu} \right. \nonumber \\ 
&& \left. -2 \left(2-\frac{\g}{2} \right) \left[\bar{\vartheta}_\mu^\mu(R,R)+\frac{D^2}{2} \la_2 \left(R+\frac{\la_2}{\la_1}D \right)  \right]  \right\}.
\ea


\section{Bounce in \texorpdfstring{$R$}{R} and \texorpdfstring{$R+cR^2$}{R+cR2} gravity and energy conditions}\label{appD}

Here we give a very brief overview of the violation of the null and weak energy conditions in $R$ (Einstein) and $R+cR^2$ (Starobinsky) gravity on a flat FLRW background. 


\subsection{Einstein gravity}

In that case of the Einstein--Hilbert action
\be
S =\frac{1}{2 \k^2} \int \rmd^4 x\, \sqrt{|g|}\, R +S_{\rm m}\, ,
\ee
where $S_{\rm m}$ is the matter action, the Friedmann equations (considering flat FLRW) are
\be
H^2 = \frac{\k^2}{3} \rho \, , \qquad
\dot{H} = -\frac{\k^2}{2} (\rho+p) \, .\label{friedR2}
\ee
Considering a bounce at $t=0$ with $H(0) = 0$ and $\dot{H}(0^+) > 0$, the second equation in \eqref{friedR2} requires 
\be
p(0)+\rho(0) < 0 \, ,
\ee
which violates the NEC \Eq{nec} and, hence, also the WEC \Eq{wec}. Therefore both the NEC and the WEC are incompatible with a cosmological bounce in Einstein gravity.


\subsection{Starobinsky gravity}\label{appD2}
From the action
\be
S =\frac{1}{2 \k^2} \int \rmd^4 x\, \sqrt{|g|}\,(R+cR^2) +S_{\rm m}\,,
\ee
the equations of motion are \cite{DeFelice:2010aj}
\be
(1+2cR)R_{\mu \nu}-\frac{1}{2}(R+c R^2)g_{\mu \nu}-2 c \N_\mu \N_\nu R+2 c g_{\mu \nu} \B R = \k^2 T_{\mu \nu} \, . \label{staroeom}
\ee
The trace of \eqref{staroeom} is
\be
\B R = \frac{1}{6c} R+\frac{\k^2}{6c} T \, , \label{tracestaro}
\ee
so that the \emph{Ansatz} \Eq{ansatzR} holds in particular for $\la_2=0$ and a radiation content (zero trace $T$), in which case $\la_1 = (6c)^{-1}$.

In the flat FLRW case, one obtains the modified Friedmann equations
\ba
\k^2 \rho &=&  3 (1+2cR)H^2-\frac{c}{2}R^2+6 c H \dot{R} \, , \label{friedRR1}  \\
\k^2 (\rho+p) &=& -2c \ddot{R}+2c \dot{R}H-2(1+2cR)\dot{H} \, .
\label{friedRR2}
\ea
From \Eqq{friedRR1}, at a bounce $H(0) = 0$ one has
\be
\k^2 \rho(0) = -\frac{c}{2}R^2 \, . \label{starorho0}
\ee
Therefore, in the case $c>0$ (corresponding to a stable scalaron in the Einstein frame), a bounce implies $\rho(0) <0$, thus violating the WEC. However, the NEC is not necessarily violated in that case. To show this (and to link this discussion with the main text), let us consider the series expansion in \eqref{Hseries} only up to $O(t^3)$ and fix $\lst=1$ and $a_1=1$, $H(t) = t+a_3 t^3+O(t^5)$. Substituting this into \eqref{friedRR2},
\be
\k^2(p+\rho)(0) = -2 \left[1+36(1+a_3)c \right]\,.
\ee
Therefore, for $c>0$ one can avoid the NEC violation if
\be
a_3 \leq -1-\frac{1}{36c}\,. \label{a3cond}
\ee
In section~\ref{genbou}, we show some cases where the fractional theory \eqref{gravac} can satisfy the WEC condition.

\end{document}